\documentclass[show packs,preprint]{revtex4}
\usepackage{graphicx}
\usepackage{dcolumn}
\usepackage{amsmath}
\usepackage{color}
\usepackage{graphics,epsfig}
\usepackage{hyperref}
\makeatletter
\def\btt#1{\texttt{\@backslashchar#1}}
\DeclareRobustCommand\bblash{\btt{\@backslashchar}} \makeatother

\begin{document}
\title{$D$-dimensional Bardeen-AdS black holes in Einstein-Gauss-Bonnet theory}
\author{Arun Kumar$^{a,}$}
\email{arunbidhan@gmail.com}
\author{Dharm Veer Singh$^{a,}$}
\email{veerdsingh@gmail.com}
 \author{Sushant G. Ghosh$^{a,b,c,}$}
\email{sgghosh@gmail.com, sghosh2@jmi.ac.in}
\affiliation{$^a$ Centre for Theoretical Physics,
 Jamia Millia Islamia,  New Delhi 110025
 India}
\affiliation{$^b$ Multidisciplinary
Center for Advanced Research and Studies(MCARS),
 Jamia Millia Islamia,  New Delhi 110025
 India}
\affiliation{$^c$ Astrophysics and Cosmology Research Unit,
 School of Mathematics, Statistics and Computer Science,
 University of KwaZulu-Natal, Private Bag X54001,
 Durban 4000, South Africa}
 
\begin{abstract}
We present a $D$-dimensional Bardeen like Anti-de Sitter (AdS) black hole solution in Einstein-Gauss-Bonnet (EGB) gravity, \textit{viz}., Bardeen-EGB-AdS black holes. The Bardeen-EGB-AdS black hole has an additional parameter due to charge ($e$), apart from mass ($M$) and Gauss-Bonnet parameter ($\alpha$). Interestingly, for each value of $\alpha$, there exist a critical $e = e_E$ which corresponds to an extremal regular black hole with degenerate horizons, while for $e< e_E$, it describes non-extremal black hole with two horizons.

Despite the complicated solution, the thermodynamical quantities, like temperature ($T$), specific heat($C$) and entropy ($S$) associated with the black hole are obtained exactly. It turns out that the heat capacity diverges at critical horizon radius $r_+ = r_C$, where the temperature attains maximum value and the Hawking-Page transition is achievable. Thus, we have an exact $D$-dimensional regular black holes, when evaporates lead to a thermodynamical stable remnant.
\end{abstract}
\maketitle
\section{INTRODUCTION}
The celebrated singularity theorems of Hawking and Penrose \cite{1} have shown that under fairly general conditions, a sufficiently massive collapsing object will undergo continual gravitational collapse, resulting in the formation of a curvature singularity.  However, the singularity is not visible to a far-away observer which essentially means that a black hole has formed. It is widely believed that these singularities do not exist in Nature, but that they are the artefact of classical general relativity. The existence of a singularity means spacetime ceases to exist, signal a breakdown of physics laws and that they must be resolved in a theory of quantum gravity \cite{rp}. While we are far from a  definite quantum gravity, attention has been shifted to regular models that are motivated by quantum arguments. The earliest idea of Sakharov \cite{Sakharov:1966} and Gliner \cite{Gliner:1966}, suggests that singularities could be avoided by matter, i.e., with a de Sitter core, with the equation of state $p=-\rho$ obeyed by the cosmological constant.

 Bardeen \cite{Bardeen:1968}, realized the idea, proposed the first regular black holes, i.e.,  there are horizons but there is no singularity.  The spherically symmetric Bardeen metric  is given by
 \begin{equation}
 \mathrm{d}s^2=g_{\mu \nu} \otimes \mathrm{d}x^{\mu} \otimes \mathrm{d}x^{\nu},\;\;\;\;(\mu,\nu=0,1,2,3),
 \end{equation}
 with $g_{\mu \nu}= \text{diag} (-f(r),f(r)^{-1},r^2,r^2\sin^2 \theta)$ and 
\begin{eqnarray}
f(r)&=&1-\frac{2mr^2}{(r^2+e^2)^{3/2}}\nonumber \\
&=&1-\left(\frac{m}{e}\right)\frac{2(r/e)^2}{(1+(r/e)^2)^{3/2}},\;\;\;\;
\text{and}\;\;\;\; r\geq0.\nonumber
\end{eqnarray}
An analysis of $f(r)=0$ reveals a critical value  $\psi^{*}$ such that $f(r)$ has a double root if $\psi=\psi^{*}$, two roots  if $\psi<\psi^{*}$ and no root if $\psi>\psi^{*}$, with $\psi=m/e$ \cite{ansoldi}. These cases illustrate, respectively, an extreme black hole with degenerate horizons, a black hole with Cauchy and event horizons, and no black hole.
Later, Ayon-Beato and Garcia \cite{ABG99} invoked nonlinear electrodynamics a to generate the Bardeen model \cite{Bardeen:1968} as an exact nonlinear magnetic monopole, i.e., an exact solution of general relativity coupled to nonlinear electrodynamics.  
The Bardeen solution is regular everywhere, that can be realized from behaviour of the scalar invariants,  Ricci scalar ($R$), Ricci square ($\mathcal{R} = R_{\mu\nu} R^{ˆ\mu\nu}$) and Kretschmann scalar ($\mathcal{K} = R_{\mu\nu\rho\sigma} R^{\mu\nu\rho\sigma}$), which are given by 
\begin{eqnarray}
R&=&\frac{6me^2(4e^2-r^2)}{(r^2+e^2)^{7/2}},\nonumber \\
\mathcal{R}&=&\frac{18m^2e^4(8e^4-4e^2r^2+13r^4)}{(r^2+e^2)^7}\nonumber\\
\mathcal{K}&=&\frac{12m^2}{(r^2+e^2)^7}\Big[8e^8-4e^6r^2+47e^4r^4-12e^2r^6+48r^8\Big],
\end{eqnarray}
where$R_{\mu,\nu}$ and $R_{\mu\nu\rho\sigma}$ are, respectively, Ricci and Reimann tensors. It is evident, that these are well behaved for $m$ and $e \neq 0$.

Subsequently, also there has been intense activities in the investigation of regular black holes and more recently \cite{8}. But most of these solutions are more or less based on Bardeen's model \cite{Bardeen:1968}. Also, some solutions in which generalized Bardeen model have been obtained in later years, which includes the Bardeen de Sitter solution \cite{Fernando:2016ksb}, rotating or Kerr-like Bardeen's solution \cite{Bambi} and noncommutative Bardeen solution \cite{Sharif:2011ja}.  

The Bardeen's regular metric  is commonly used to compare with the classical black hole, in various applications which include thermodynamical properties \cite{12}, geodesics equations \cite{Stuchlik:2014qja}, quasinormal modes \cite{14}, Hawking evaporation \cite{15} and black hole's remnant \cite{Mehdipour:2016rtd}. The rotating Bardeen regular metric has been tested with a black hole candidate in Cygnus X-1 \cite{Bambi:2014nta}, and also shown that the it can also act as a natural particle accelerator \cite{Ghosh:2015pra}. Lately, the Bardeen's solution is extended to higher dimensional spacetime \cite{sabir}. 

Last few decades gravity witnessed considerable activities in higher dimensions motivated by the superstring and field theories.  In addition to higher-curvature corrections to Einstein theory, string theory makes several predictions about nature, the most important ones are the existence of extra dimensions \cite{gross}. The Einstein-Gauss-Bonnet gravity is a  natural and most effective generalization of Einstein’s general relativity, to higher dimensions, motivated by the heterotic string theory. It was discovered first by Lanczos \cite{5}, and rediscovered by David Lovelock \cite{6}. The Einstein-Gauss-Bonnet (EGB) theory allow us to explore several conceptual issues of gravity in a broader setup and the theory is known to be free of ghosts while expanding about the flat space \cite{BD}. The effective field equations, in the EGB theory,  are of second-order like in general relativity, but admit, in $D$ $> 5$, new black hole solutions \cite{cai} that are unavailable to the Einstein theory. The first black hole solutions of EGB theory was obtained by Boulware and Deser \cite{BD} which is similar to its general relativity counterpart with a curvature singularity at $r=0$.   Later several authors studied exact black hole solutions in EGB theory and their thermodynamical properties  \cite{26,sus}.   

The black holes with higher derivative curvature in Anti-de Sitter (AdS) spaces have been considered in the recent years, e.g., static AdS black hole solutions in EGB gravity with several interesting features \cite{27,Neu}. It is the purpose of this paper to obtain a $D$-dimensional spherically symmetric Bardeen-like black holes solution for the EGB gravity in AdS spacetimes, viz., EGB-Bardeen-AdS metric. It is shown that the EGB-Bardeen-AdS metric is an exact black hole solution of EGB  coupled to nonlinear electrodynamics in AdS spacetime thereby generalizing the Boulware-Desser solution \cite{BD} which is encompassed as a special case.  We analyze their thermodynamical properties to find a stable black hole remnant and also perform a thermodynamic stability analysis of the EGB-Bardeen-AdS black holes.

The paper is ordered as: we obtain $D$-dimensional EGB-Bardeen-AdS black hole metric for a nonlinear electrodynamics as a source in Sec. II and also give the basic equations governing EGB theory. We investigate the structure and location of the horizons of the $D$-dimensional EGB Bardeen Black holes metric along with their energy conditions in Sec. II. Sec. III is devoted to the study of the thermodynamical properties of $D$-dimensional EGB Bardeen Black holes with a focus on the stability and also discuss black hole's remnant. We end the paper with our concluding remarks in Sec. V. We use the units such that $G=c=1$.

\section{EINSTEIN-GAUSS-BONNET  with NONLINEAR ELECTRODYNAMICS}
Our paper begins with the action of Einstein-Gauss-Bonnet gravity with the negative cosmological constant coupled to nonlinear electrodynamics \cite{MS1} which reads:
\begin{equation}
\mathcal{I}_{G}=\frac{1}{2}\int_{\mathcal{M}}d^{D}x\sqrt{-g}\left[  R +\alpha\mathcal{L}_{GB}+\frac{(D-1)(D-2)}{l^2} +\mathcal{L}(F) \right]
\end{equation}
where $\Lambda=-(D-1)(D-2)/2l^2$ is the cosmological constant and $\alpha$ is the Gauss-Bonnet coupling coefficient with dimension [length]$^2$. The discussion will be given here corresponding to the case with $\alpha\geq0$ \cite{djg,bento}. The nonlinear electrodynamics is described by $\mathcal{L}(F)$ in the invariant $ {F}=F_{\mu\nu}F^{\mu\nu}/4$,  where $F_{\mu\nu}$ is associated with gauge potential $A_{\nu}$ via $F_{\mu\nu}=2\nabla_{[\mu}A_{\nu]}$. The Gauss-Bonnet Lagrangian is of the form \cite{cai,sus,MS1}
\begin{equation}
\mathcal{L}_{GB}=R_{\mu\nu\gamma\delta}R^{\mu\nu\gamma\delta}-4R_{\mu\nu}R^{\mu\nu}+R^{2},
\end{equation}
 Here $R_{\mu\nu}$, $R_{\mu\nu\gamma\delta}$ and $R$ are respectively the Ricci tensors, Riemann tensors, and  Ricci scalar. The variation of the action with respect to the metric $g_{\mu\nu}$ gives the following EGB equations of motion \cite{sus}%
\begin{eqnarray}\label{ee}
&&G_{\mu\nu}+\alpha H_{\mu\nu}+\Lambda g_{\mu\nu} = T_{\mu\nu}\equiv 2\left[\frac{\partial \mathcal{L}(F)}{\partial F}F_{\mu\rho}F_{\nu}^{\rho}-g_{\mu\nu}\mathcal{L}(F)\right],
\end{eqnarray}
\begin{eqnarray}\label{ee1}
&&\nabla_{\mu}\left(\frac{\partial \mathcal{L(F)}}{\partial F}F_{\mu\nu}\right)=0~~~ \text{and}~~~\nabla_{\mu}\left(^*F_{\mu\nu}\right)=0.
\end{eqnarray}
where
$G_{\mu\nu}$ is the Einstein tensor and $H_{\mu\nu}$ is Lanczos tensor \cite{Hendi1}
\begin{eqnarray}
&& H_{\mu\nu}  =  2\;\Big( -R_{\mu\sigma\kappa\tau}R_{\quad\nu}^{\kappa
\tau\sigma}-2R_{\mu\rho\nu\sigma}R^{\rho\sigma}-2R_{\mu\sigma}R_{\ \nu
}^{\sigma}+RR_{\mu\nu}\Big) -\frac{1}{2}\mathcal{L} _{GB}g_{\mu\nu},
\end{eqnarray}
Following \cite{Fernando:2016ksb}, we have the Lagrangian density of the matter field is \cite{sabir}
\begin{eqnarray}
\mathcal{L}(F) =\frac{1}{4se^2}\left[\frac{\sqrt{2e^2F}}{1+\sqrt{2e^2F}} \right]^\frac{2D-3}{D-2},
\end{eqnarray}
with
 \begin{equation}
s = \begin{cases}

     \frac{e^{D-3}}{(D-1)\mu'^{D-3}},& D=\text{even},\\

   \frac{e^{D-3}}{(D-3)\mu'^{D-3}}, & D=\text{odd}.\\
 \end{cases}
 \end{equation}

 We consider the following \textit{anstaz} for the Maxwell field \cite{sabir}
\begin{eqnarray}
F_{\mu\nu}&=&2\delta^{\theta_{1}}_{[\mu}\delta^{\theta_{2}}_{\nu]}g\sin\theta_{1};\qquad\qquad\qquad\qquad\qquad \qquad\qquad D=4,\nonumber\\
F_{\mu\nu}&=&2\delta^{\theta_{D-3}}_{[\mu}\delta^{\theta_{D-2}}_{\nu]}\frac{g^{D-3}}{r^{D-4}}\sin\theta_{D-3}\left[\prod_{j=1}^{D-4}\sin^2\theta_{j}\right];\qquad\quad\,  D\geq 5.
\end{eqnarray}
Eq. (\ref{ee1}) implies that $dF=0$, thereby we obtains
\begin{equation}
e'(r)2\delta^{\theta_{D-3}}_{[\mu}\delta^{\theta_{D-2}}_{\nu]}\frac{g^{D-3}}{r^{D-4}}\sin\theta_{D-3}\left[\prod_{j=1}^{D-4}\sin^2\theta_{j}\right] d\theta\wedge d\phi\wedge\hdots\wedge d\psi_{(D-2)}.
\end{equation}
This leads to $e(r)=e=$ constant. Interestingly, the other components of $F_{\mu\nu}$ have negligible influence in comparison of $F_{\theta\phi}$ \cite{sabir, Hendi1}. The energy momentum tensor can be given as
\begin{eqnarray}\label{em}
T^t_t &=& T^r_r =\beta\frac{(D-2)\mu'e^{D-2}}{(r^{D-3}+e^{D-3})^{\frac{2D-3}{D-2}}},
\end{eqnarray}with
\begin{equation}
 \beta= \begin{cases} 
 (D-1),& D=\text{even},\\
   (D-4), & D=\text{odd}.\\
 \end{cases}
 \end{equation}
 Using energy momentum tensor from Eq. (\ref{em}), we can obtain the Bardeen-EGB-Ads black hole solution in the following section.
\section{Bardeen Anti de-Sitter Black Holes in EGB theory}
We wish to obtain $D$-dimensional static spherically symmetric solutions of Eq.~(\ref{ee}). We assume the metric to be of the following form \cite{sus,Hendi1}
\begin{equation}\label{metric}
ds^2 = -f(r) dt^2+ \frac{1}{f(r)} dr^2 + r^2 \tilde{\gamma}_{ij}\; dx^i\; dx^j,
\end{equation}
where $ \tilde{\gamma}_{ij} $ is the metric of a $(D-2)$-dimensional constant curvature space $k = 1,\; 0,\;$ or -1. The spherically symmetric static black hole solution of EGB theory was obtained by Boulware and Deser \cite{BD}.
Using metric (\ref{metric}), the $(r,r)$ equation of field equation reduces to %
\begin{eqnarray}
& & (D-2)\Big[\left( r^{3}-2\tilde{\alpha}r\left(  f\left(  r\right)  -1\right)
\right)  f^{\prime}\left(  r\right)  +  \left(  D-3\right)  r^{2}\left(  f\left(  r\right)  -1\right)\nonumber\\  &&~~~~\qquad\qquad\qquad\qquad-\left(
D-5\right)  \tilde{\alpha}\left(  f\left(  r\right)  -1\right)
^{2}\Big]+\Lambda  =\beta\frac{\mu'e^{D-2}}{(r^{D-3}+e^{D-3})^{\frac{2D-3}{D-2}}},
\label{eom1}
\end{eqnarray} 
 where  prime denotes a derivative with respect to $r$ and  $\tilde{\alpha}=$ $\left(  D-3\right)  \left(  D-4\right)  \alpha$.
 The Eq. (\ref{eom1}) can be easily integrated to give general solution as
\begin{equation} \label{sol:egb}
f_{\pm}\left(  r\right)  =1+\frac{r^{2}}{2\tilde{\alpha}}\left(  1\pm
\sqrt{1+\frac{4\tilde{\alpha}\mu'}{\left(r^{D-2}+e^{D-2}\right)^{\frac{D-1}{D-2}}}-\frac{4\tilde{\alpha}}{l^2}}\right)  ,\qquad\qquad\text{ \ }D\geq5
\end{equation}
where $\mu'$ is the mass of the black hole it is related to the Arnowitt-Deser-Misner (ADM) mass $M$ with relation \cite{sus}
\begin{equation}\label{a2}
\mu' =\frac{16 \pi M}{(D-2) V_{D-2}},\qquad \text{with}\qquad V_{D-2}=\frac{2\pi^{(D-1)/2}}{\Gamma{(D-1)/2}},
\end{equation}  
where $V_{D-2}$ is the volume of the $(D-2)$-dimensional unit sphere. There are two families of solutions which correspond to the signs $(\pm )$ in front of square root in (\ref{sol:egb}). The solution (14) with (16) is a general spherically symmetric $D$-dimensional solution of EGB theory coupled to nonlinear electrodynamics in an AdS spacetime thereby generalizing the Bardeen solution.  The special case in which charge $e=0$ and $\Lambda=0$, one get the Boulware-Deser solution \cite{BD}. It see that solution (14) with (16) gets for other field equation. For definiteness, henceforth, we shall call solution (\ref{sol:egb}) Bardeen-EGB-AdS black holes. In the  case of no charge $e = 0$, Eq. (\ref{sol:egb}) reduces to $D$-dimensional EGB-AdS black holes \cite{Neu,Ads,MS1} and in the limit, $\alpha\to 0$, the negative branch of (\ref{sol:egb}) to $D$- dimensional Bardeen-AdS black holes \cite{sabir} 
\begin{equation}\label{sabir}
ds^2 = -\left[1-\frac{\mu'r^2}{(r^{D-2}+e^{D-2})^{\frac{D-1}{D-2}}}+\frac{r^2}{l^2}\right] dt^2+ \frac{1}{\left[1-\frac{\mu'r^2}{(r^{D-2}+e^{D-2})^{\frac{D-1}{D-2}}}+\frac{r^2}{l^2} \right]} dr^2 + r^2 \tilde{\gamma}_{ij}\; dx^i\; dx^j. 
\end{equation}
Further, the solution also goes over to Schwarzschild-Tangherlini black hole \cite{st}in the absence of charge. To study the structure of solution, we take limit $r\to 0$ to obtain
\begin{eqnarray}
&&f(r)=1+\frac{r^2}{l^2_{{eff}}};\qquad\qquad\quad r\to 0,
\end{eqnarray}
where  ($1/l_{eff}^2$) is effective AdS length, it reads 
 \begin{equation}
\frac{1}{l_{{eff}}^2}=\frac{1}{2\tilde{\alpha}}\left(1-\sqrt{1+\frac{4\mu'\tilde{\alpha}}{e^{D-1}}-\frac{4\tilde{\alpha}}{l^2}}\right),
\end{equation}
which is describing a de Sitter solution for $\tilde{\alpha}>0$ in the Bardeen-EGB-AdS black hole.

The regularity of the black hole solution (\ref{sol:egb}) can be seen by behaviour of the scalar invariants, which are given by 
\begin{eqnarray}\label{inv}
&&\lim_{r\rightarrow0} R = \frac{D(D-1)}{2\tilde{\alpha}}\left[-1+\left(1+\frac{4\mu'\tilde{\alpha}}{e^{D-1}}-\frac{4\tilde{\alpha}}{l^2}\right)^{1/2}\right],\nonumber\\
&&\lim_{r\rightarrow0} \mathcal{R} = \frac{D(D-1)^2}{2\tilde{\alpha}^2}\left[1+\frac{2\mu'\tilde{\alpha}}{e^{D-1}}-\frac{2\tilde{\alpha}}{l^2}-\left(1+\frac{4\mu'\tilde{\alpha}}{e^{D-1}}-\frac{4\tilde{\alpha}}{l^2}\right)^{1/2}\right],\nonumber\\
&&\lim_{r\rightarrow0}\mathcal{K} = \frac{D(D-1)}{\tilde{\alpha}^2}\left[1+\frac{2\mu'\tilde{\alpha}}{e^{D-1}}-\frac{2\tilde{\alpha}}{l^2}-\left(1+\frac{4\mu'\tilde{\alpha}}{e^{D-1}}-\frac{4\tilde{\alpha}}{l^2}\right)^{1/2}\right].
\end{eqnarray}
 Thus, the spacetime is regular everywhere as seen from the behavior of other invariants if $\mu'=e=0$. The $D$-dimensional Bardeen-EGB-AdS black hole solution is well defined everywhere by its curvature invariants.

\begin{figure*} 
\begin{tabular}{c c c c}
\includegraphics[width=0.5\linewidth]{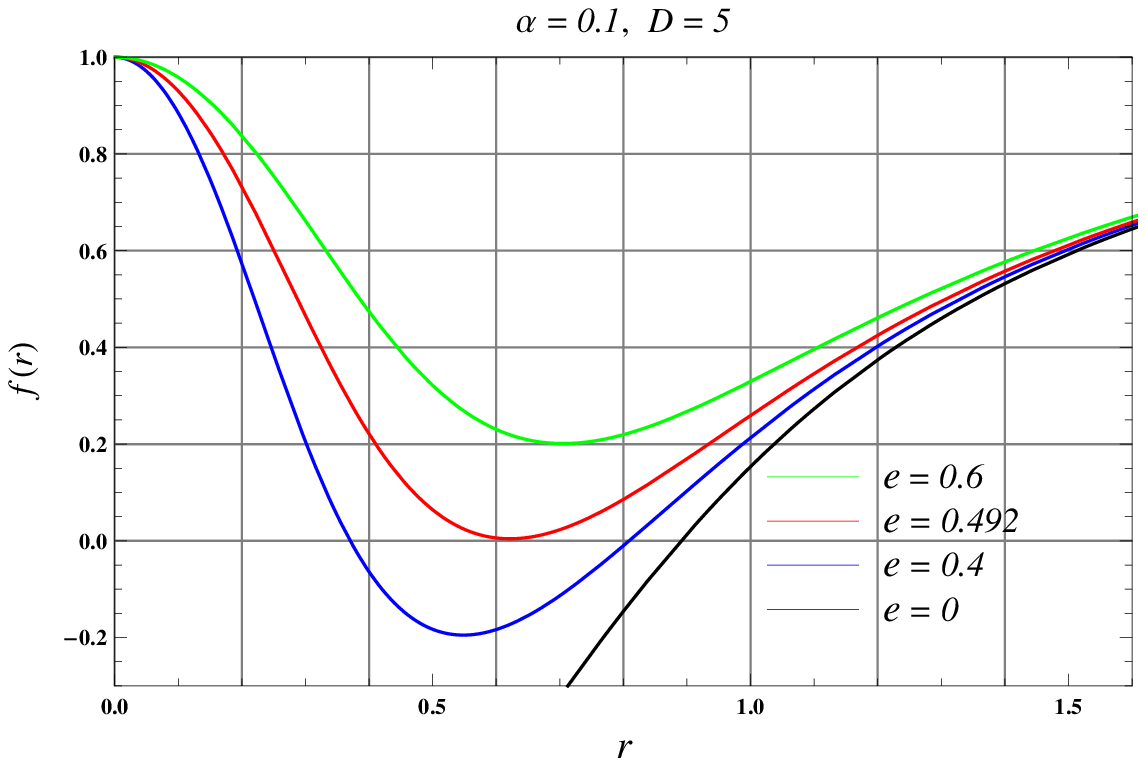}
\includegraphics[width=0.5\linewidth]{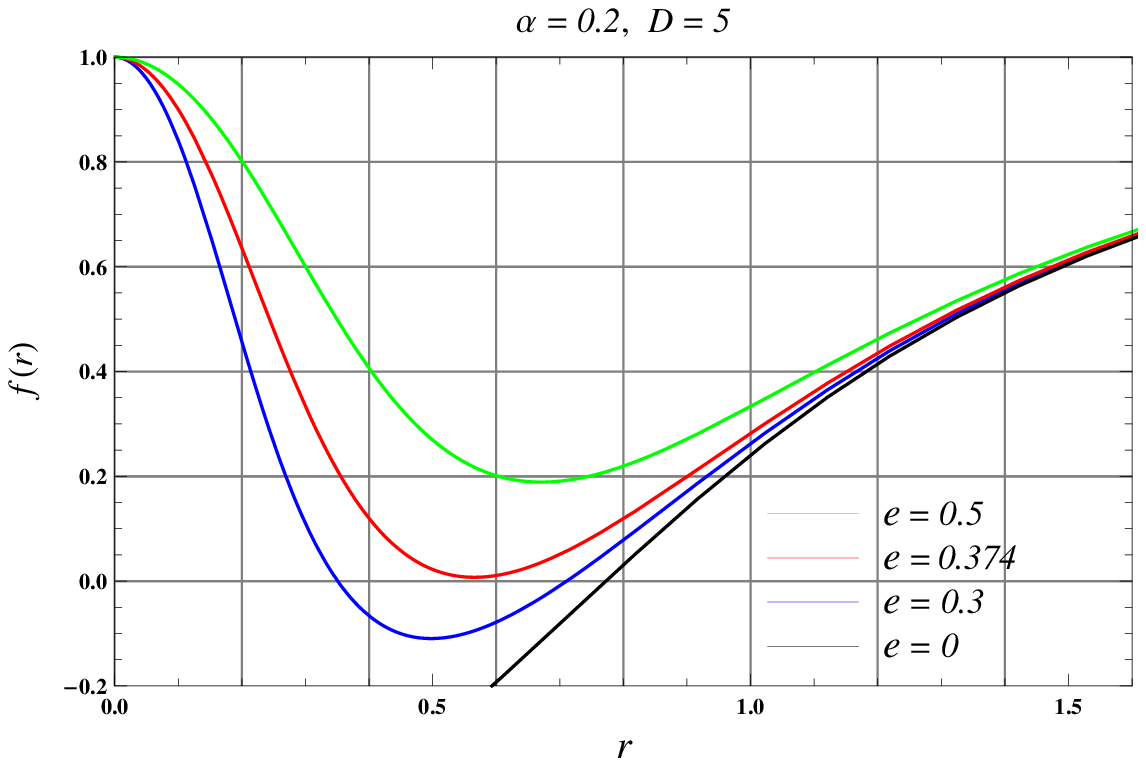}\\
\includegraphics[width=0.5\linewidth]{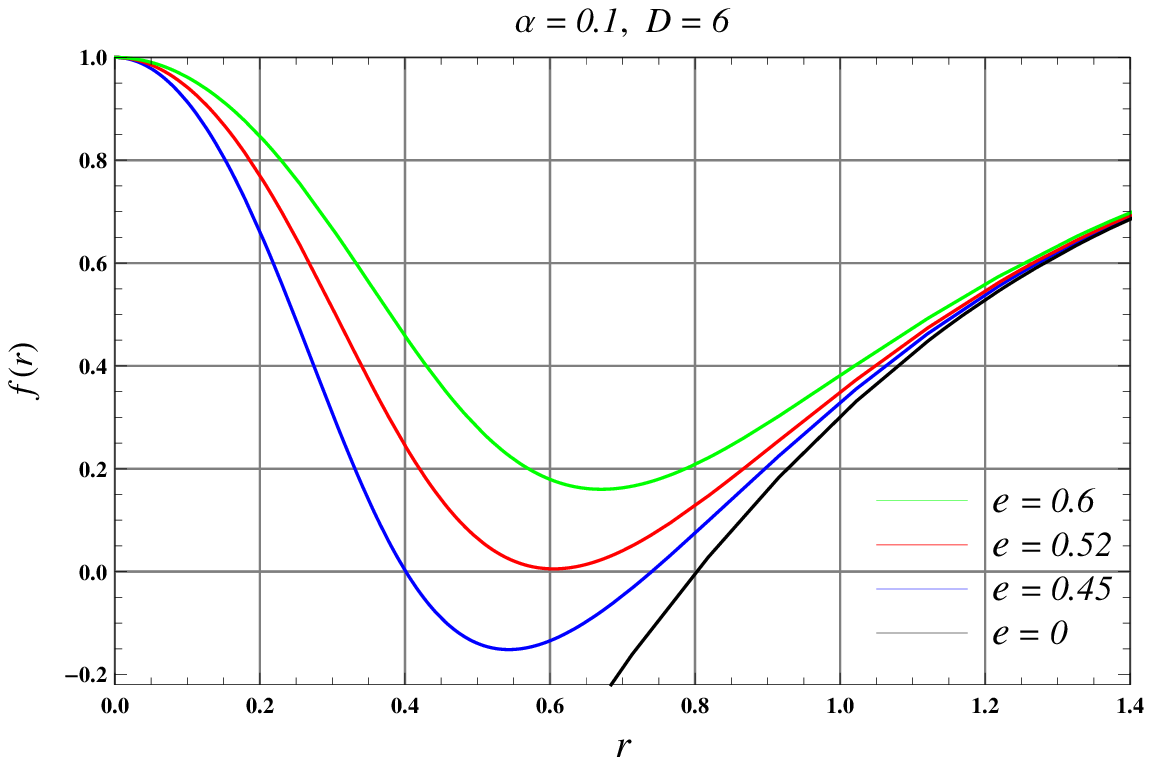}
\includegraphics[width=0.5\linewidth]{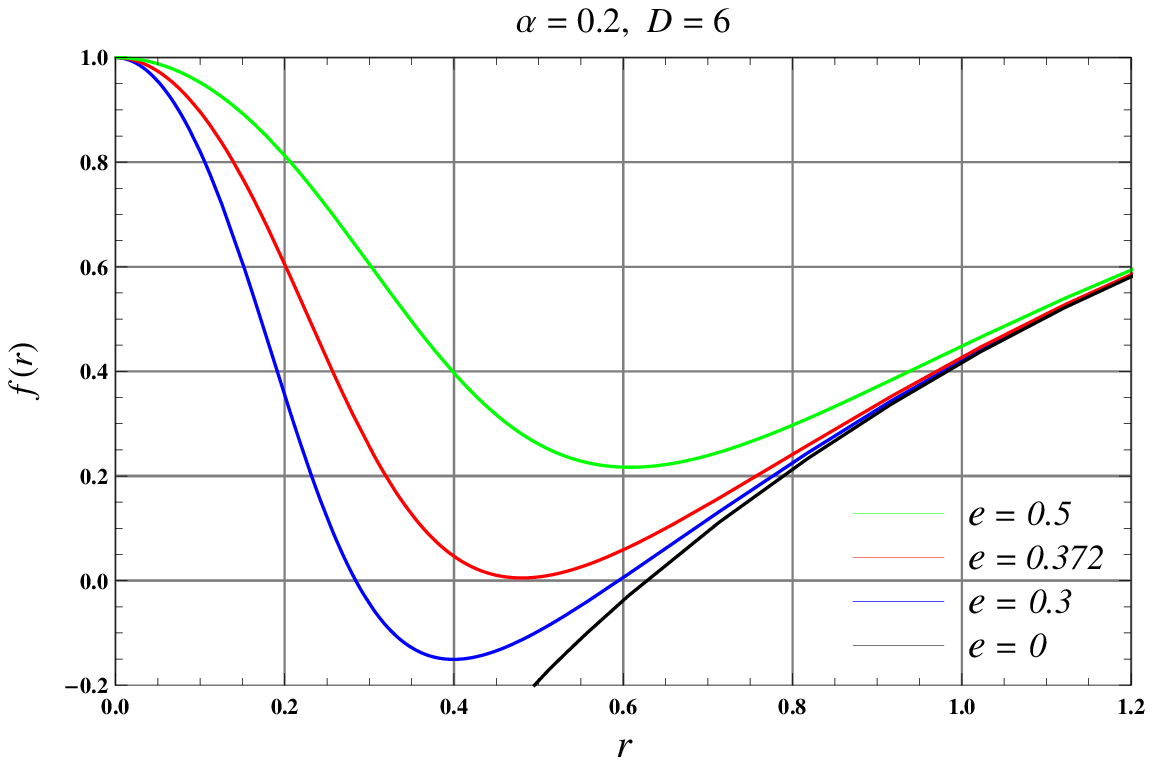}\\
\includegraphics[width=0.5\linewidth]{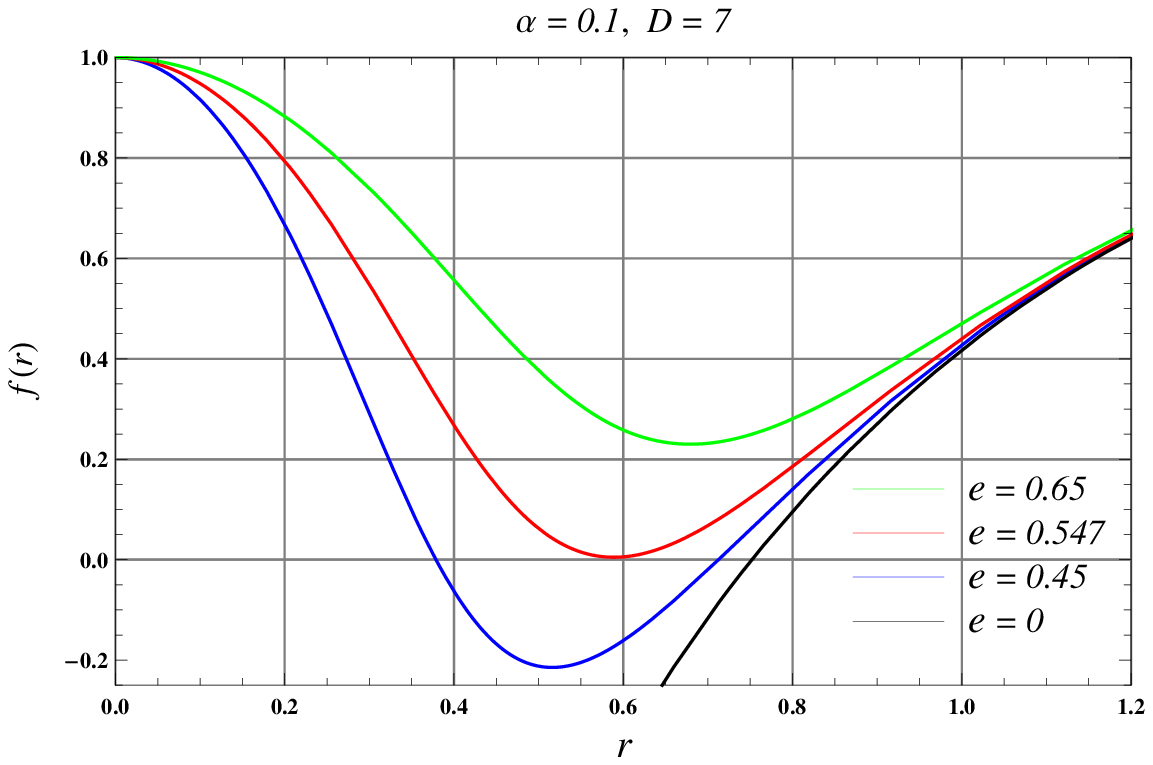}
\includegraphics[width=0.5\linewidth]{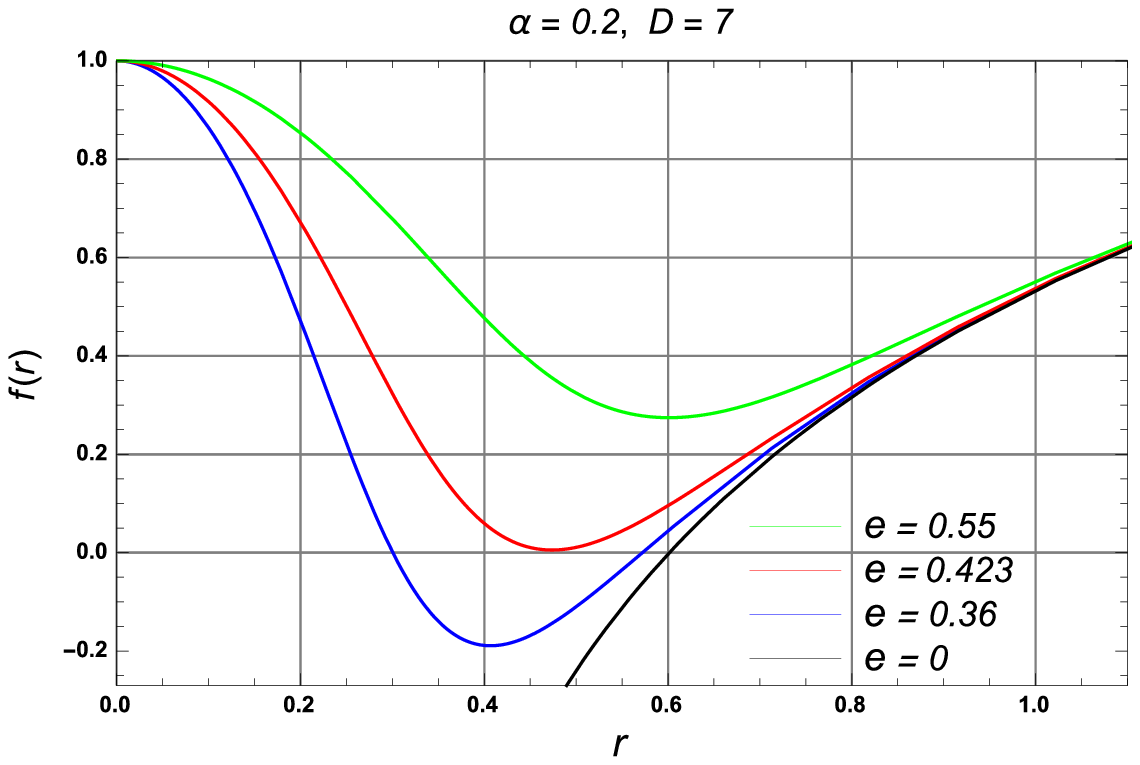}\\
\includegraphics[width=0.5\linewidth]{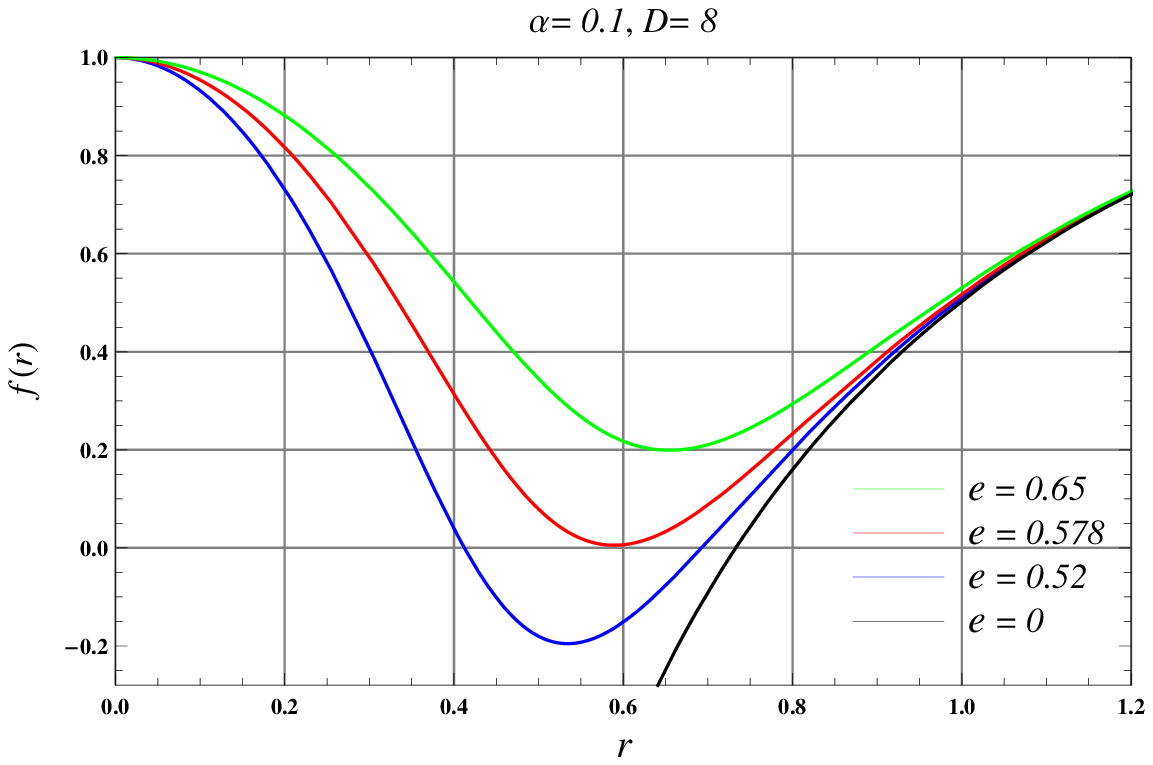}
\includegraphics[width=0.5\linewidth]{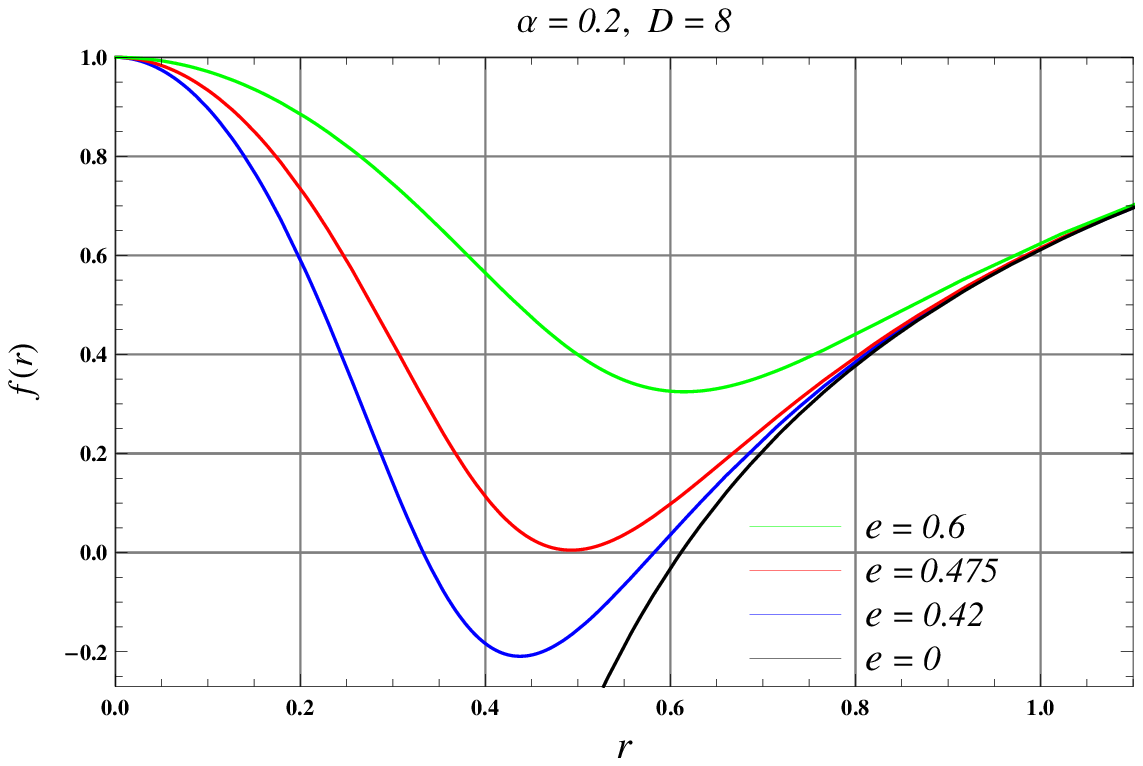}
\end{tabular}
\caption{\label{fig:th} Plot of metric function $f(r)$ vs $r$  in various dimensions $D$ = 5, 6, 7, and 8 (top to bottom) for different values of charge $e$ with Gauss-Bonnet coupling  $\alpha$ = 0.1 and 0.2 (left to right) with $\mu'=1$. }
\end{figure*}
 The weak energy condition states that $T_{ab}t^at^b\geq 0$ for all time like vectors $t^a$, i.e., the local energy density cannot be negative for any observer. The dominant energy condition states that $T_{ab}t^at^b\geq 0$  and $T^{ab}t_b$ must be space like, for any time like vector $t^a$. 
Hence, the energy conditions require $\rho\geq 0$ and $\rho+P_i\geq 0$,
\begin{eqnarray}
\rho+P_2=\rho+P_3=\rho+P_4=\beta\frac{(D-2)\mu'e^{D-2}}{(r^{D-3}+e^{D-3})^{\frac{2D-3}{D-2}}}.
\end{eqnarray}
Where $\beta=(D-1)$ and $(D-4)$ are, respectively, for even and odd dimensions.  It is worthwhile to note that  $\rho > P_3 $  and $P_1 = - \rho$. Thus the Bardeen-EGB-AdS black holes obey the weak energy condition. 

Next, we proceed to discuss the horizon structure of our Bardeen-EGB-AdS black holes. The horizons radius, if exists, are zeros of $g^{rr}=f(r)=0$.
 The numerical analysis of $f(r)=0$ reveals that  it is possible to find non-vanishing value of $\alpha$ and $e$ for which metric function $f(r)$ is minimum, i.e, $f(r)=0$ admits two roots $r_{\pm}$. The smaller and larger roots, respectively, corresponds to the Cauchy and event horizon of the black holes. We have shown that for a given value of $\alpha$ and fixed $\mu'$, there exists a critical charge parameter $e_E$, and  critical horizon radius $r_E$, such that $f(r_E)=0$ has a double root, i.e, $r_E=r_{\pm}$. This case corresponds to the extremal Bardeen-EGB-AdS black holes with  degenerate horizons. When $e < e_E$  the two horizons $r_{\pm}$ correspond to the non-extremal black hole and $e>e_E$ has no horizon, i.e., no black holes (cf. Fig. \ref{fig:th} and Table 1). It is clear that the critical value of $e_E$ and $r_E$ depend upon the coupling constant $\alpha$. For $\alpha=0.1$, $0.2$ the  critical value of the charge corresponds to the degenerate horizon for $D=5,6,7 $ and $8$ are shown in Table \ref{tahhor1}. Also, the radius of the event horizon decreases  with increase in Gauss-Bonnet coefficient $\alpha$ and increases with charge $e$ and dimensions $D$ as shown in Fig. \ref{fig:th}.

\begin{table}[h]
    \begin{center}        
        \begin{tabular}{|l| l l l l  | l l l l  |  }
            \hline
         \multicolumn{1}{|c}{}  & \multicolumn{4}{|c|}{$\alpha=0.1$}  &\multicolumn{4}{c|}{$\alpha=0.2$}\\
            \hline
           \,\, \,\,Dimensions~~&~~$e$ & ~~~$r_-$ &~~~$r_+$ &~~~~ $\delta$ &~~ $e$ &~~~~$r_-$  &~~~~ $r_+$ &~~~~$\delta$  \\
            \hline
             \,\,\,\,&\,\,$e_E$=0.493 & ~0.6242  & ~~0.6242 &~~ 0    \,\,   &$e_E=$0.373 &~ 0.566& ~~0.566&~   0         \\
            \,\,\,\,$D=5$&\,\,0.3 & ~0.2519  & ~~0.8695 &~~ 0.6176 \,\,   &~~0.2   &~ 0.2152& ~~0.7632&~  0.548        \\
            \,\,\,\,&\,\,0.4 & ~0.3746  & ~~0.8164 &~~ 0.4418  \,\,  &~~0.3   &~ 0.3501& ~~0.7182&~  0.3681 \,\,         \\
            \hline
             \,\,\,\,&\,\,$e_E$=0.52 & ~0.607  & ~~0.607 &~~ 0    \,\,   &$e_E=$0.372 &~ 0.4773& ~~0.4773&~   0         \\
            \,\,\,\,$D=6$&\,\,0.3 & ~0.2619  & ~~0.8038 &~~ 0.5419 \,\,   &~~0.2   &~ 0.1583& ~~0.6276&~  0.4693 \,\,      \\
            \,\,\,\,&\,\,0.4 & ~0.3278  & ~~0.768 &~~ 0.4402  \,\, &~~0.3 \,\,  &~ 0.2871& ~~0.5969&~  0.3098\,\,          \\
            
           \hline
             \,\,\,\,&\,\,$e_E$=0.547 & ~0.5939  & ~~0.5939 &~~ 0 \,\,      &$e_E=$0.423 &~ 0.4656& ~~0.4656&~   0         \\
            \,\,\,\,$D=7$\,\,\,\,&\,\,0.3 & ~0.1767  & ~~0.7564 &~~ 0.5797 \,\,   &~~0.2   &~ 0.1142& ~~0.6062&~  0.4920\,\,       \\
            \,\,\,\,\,\,&\,\,0.4 & ~0.2840  & ~~0.7442 &~~ 0.4602 \,\,  &~~0.3   &~ 0.2182& ~~0.5950&~  0.3768 \,\,   \\
\hline
             \,\,\,\,&\,\,$e_E$=0.578 & ~0.5816  & ~~0.5816 &~~ 0\,\,       &$e_E=$0.475 &~ 0.4881& ~~0.4881&~   0         \\
            \,\,\,\,$D=8$&\,\,0.3 & ~0.1460  & ~~0.7380 &~~ 0.5920 \,\,   &~~0.2   &~ 0.0860& ~~0.6175&~  0.5315  \,\,    \\
            \,\,\,\,&\,\,0.4 & ~0.2442  & ~~0.7319 &~~ 0.4877   &~~0.3\,\,   &~ 0.1732& ~~0.6090&~  0.4358  \,\,  \\
            \hline 
        \end{tabular}
        \caption{Radius of Cauchy horizon ($r_-$), the event horizon ($r_+$) and $\delta = r_+-r_-$ for different values of charge $e$ and dimension $D$.}
\label{tahhor1}
    \end{center}
\end{table}
 \section{Black Hole Thermodynamics}

 In this section, we explore the thermodynamics of the Bardeen-EGB-AdS black holes. Henceforth, we shall restrict our discussion to the negative branch of solution (\ref{sol:egb}). The black hole thermodynamics provides the insight into quantum properties of gravitational field, in particular, the thermodynamics of AdS black holes has been of great interest to the astrophysicists since the pioneering work by Hawking and Page \cite{hp}, who suggested the existence of a phase transition in AdS black holes. The Bardeen-EGB-AdS black hole is characterized by mass $M$, charge $e$ and $\Lambda$. The black hole mass can be determined by using $f(r_+)=0$ in terms of horizon radius $r_+$ as  
\begin{eqnarray}
M_+ =\frac{(D-2)V_{D-2}~r_{+}^{D-3}}{16 \pi} \left[\left(1 + \frac{\tilde{\alpha}}{r_{+}^2}+\frac{r_{+}^2}{l^2}\right)\left(1+\frac{e^{D-2}}{r_+^{D-2}}\right)^\frac{D-1}{D-2}\right]. \label{M1}
\label{mass1}
\end{eqnarray}
 The mass expression (\ref{mass1}) reduces to the mass of EGB-AdS black hole \cite{Neu,cai,Ads,MS1} in the absence of charge $(e=0)$ as
\begin{equation}
M_{+} = \frac{(D-2)  V_{D-2}~r_{+}^{D-3} }{16 \pi}   \left[1 + \frac{\tilde{\alpha}}{r_{+}^2}+\frac{r_{+}^2}{l^2}\right],
\end{equation}
  we recover the mass obtained for the EGB black hole \cite{sus,Wij} when $e=0, \text{and}\, \Lambda=0$ and further  the mass for $D$-dimensional Bardeen-AdS black hole \cite{sabir}, in the limit $\alpha \to 0$, yields 
\begin{eqnarray}
M_{+} = \frac{(D-2)  V_{D-2}~r_{+}^{D-3} }{16 \pi}  \left[\left(1+\frac{r_+^2}{l^2}\right)\left(1+\frac{e^{D-2}}{r_+^{D-2}}\right)^{\frac{D-1}{D-2}}\right].
\end{eqnarray}
The Eq. (\ref{mass1}) reduce to the mass of  Schwarzschild-Tangherlini black hole \cite{sus,PK} when $e=0$, $\alpha \to 0$ .
The black hole does have a temperature known as Hawking temperature defined by $T=\kappa/2\pi$, where $\kappa$ is the surface gravity given by \cite{sus,PK}
\begin{eqnarray}
\kappa^2=-\frac{1}{2}\nabla_\mu\xi_\nu \nabla^\mu \xi^\nu,
\end{eqnarray}
where $\xi^\mu$ is a Killing vector. For static spherically symmetric case the Killing vector  $\xi^\mu$, takes the form  $\xi^\mu=\partial^\mu_t$. Using the metric function (\ref{metric}), the surface gravity takes the following form
\begin{eqnarray}\label{sg1}
\kappa=\frac{1}{2}\frac{\partial{\sqrt{-g^{rr} g_{tt}}}}{\partial r}\mid_{r=r_+}=\frac{1}{2}\frac{df(r)}{dr}\mid_{r=r_+}.
\end{eqnarray}
Hence, using (\ref{sol:egb}), the Hawking temperature for the Bardeen-EGB-AdS black hole can be calculated as
 \begin{eqnarray}
T_+ &=& \frac{1}{4 \pi r_+} \left[\frac{(D-3)r_{+}^2+(D-5)\tilde{\alpha}-\frac{2e^{D-2}}{r_{+}^{D-2}}(r_{+}^2+2\tilde{\alpha})+\frac{D-1}{l^2}r_{+}^4}{ (r_{+}^2+2\tilde{\alpha})(1+\frac{e^{D-2}}{r_{+}^{D-2}})}\right].
 \label{temp1}
\end{eqnarray}
The Hawking temperature is positive if
\begin{equation}
 e^{D-2}<\frac{(D-3)l^2r_+^d+(D-1)r_+^{d+2}+(D-5)\tilde{ \alpha}l^2}{2l^2(r_+^2+2\tilde{ \alpha})}
\end{equation}
  Note that the charge term modifies the Hawking temperature of EGB black holes, and taking limit $e = 0, \Lambda= 0$, we recover the EGB black hole \cite{Wij,sus} temperature as
\begin{eqnarray}
T_+ &=& \frac{1}{4 \pi r_+} \left[\frac{(D-3)r_{+}^2+(D-5)\tilde{\alpha}}{ r_{+}^2+2\tilde{\alpha}}\right],
 \label{temp2}
\end{eqnarray}
 We recover the $D$-dimensional Bardeen black hole temperature when $\alpha = \Lambda=0$
\begin{eqnarray}
T_+ = \frac{1}{4 \pi r_+} \left[\frac{(D-3)-2\frac{e^{D-2}}{r_{+}^{D-2}}}{ 1+\frac{e^{D-2}}{r_+^{D-2}}}\right],
\end{eqnarray}
which further reduces to $T_+ =(D-3)/{4 \pi r_+}$ for Schwarzschild-Tangherlini black hole \cite{sus,PK} in the absence of charge ($e= 0$). The maximum temperature occurs at the critical radius $r_C^T$ shown in Table II. We can say that horizon radius $r_E$ of the extremal black hole corresponds to zero temperature $T_+=0$.
 \begin{table}[h]
    \begin{center}        
        \begin{tabular}{|l| l l l   |  l l l  |  }
            \hline
         \multicolumn{1}{|c}{}  & \multicolumn{3}{|c|}{$\alpha=0.1$}  &\multicolumn{3}{c|}{$\alpha=0.2$}\\
            \hline
           \,\, \,\,Dimensions~~&$e$ & ~~~$r_c^T$ &~~~$T_+^{Max}$  &~~ $e$ &~~~~$r_c^T$  &~~~~ $T_+^{Max}$   \\
            \hline

            \,\,\,\,$D=5$&\,\,0.4 & ~1.040  & ~~0.09785  \,\,   &~~0.3   &~ 1.088& ~~0.08127        \\
\,\,\,\,&\,\,0.493 & ~1.115  & ~~0.08723    \,\,   &\,\,0.373 &~ 1.367&~  0.0786      \\
            \,\,\,\,&\,\,0.6 & ~1.330  & ~~0.08252   \,\,  &~~0.5   &~ 1.400& ~~0.07480 \,\,         \\
            \hline

            \,\,\,\,$D=6$&\,\,0.4 & ~1.083  & ~~0.12140 \,\,   &~~0.3   &~ 0.914& ~~0.09704 \,\,      \\
\,\,\,\,&\,\,0.52 & ~1.179  & ~~0.11610     \,\,   &\,\,0.374 &~ 1.099& ~~0.09477       \\
            \,\,\,\,&\,\,0.6 & ~1.287  & ~~0.11180  \,\, &~~0.5 \,\,  &~ 1.356& ~~0.09114\,\,          \\
            
           \hline

            \,\,\,\,$D=7$\,\,\,\,&\,\,0.45 & ~0.871  & ~~0.14500 \,\,   &~~0.36   &~ 0.668& ~~0.13180\,\,       \\
\,\,\,\,&\,\,0.547 & ~1.069  & ~~0.13760  \,\,      &\,\,0.423 &~ 0.814& ~~0.12130         \\
            \,\,\,\,\,\,&\,\,0.65 & ~1.274  & ~~0.13090 \,\,  &~~0.55   &~ 1.075& ~~0.10830 \,\,   \\
\hline

            \,\,\,\,$D=8$&\,\,0.52 & ~0.869  & ~~0.17070  \,\,   &~~0.42   &~ 0.666& ~~0.17440  \,\,    \\
\,\,\,\,&\,\,0.578 & ~0.964  & ~~0.16250 \,\,       &\,\,0.475 &~ 0.772& ~~0.15880        \\
            \,\,\,\,&\,\,0.65 & ~1.105  & ~~0.15500    &~~0.6\,\,   &~ 0.983& ~~0.13630  \,\,  \\
            \hline 
        \end{tabular}
        \caption{The maximum Hawking temperature $T_+^{Max}$ at critical radius $r_C^T$ for different values of charge $e$ and different dimension $D=5,6,7$  and $8$ with fixed value of $l=10$.}
\label{tahhor}
    \end{center}
\end{table}
\begin{figure*} 
\begin{tabular}{c c c c}
\includegraphics[width=0.5\linewidth]{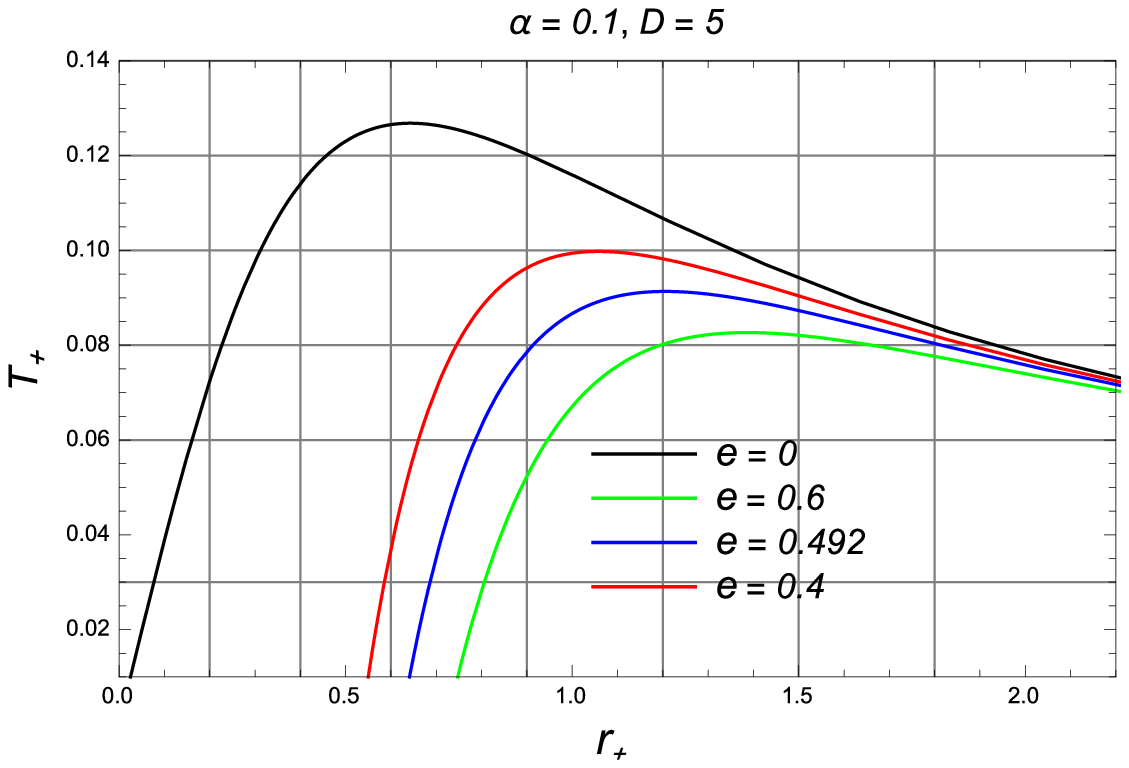}
\includegraphics[width=0.5\linewidth]{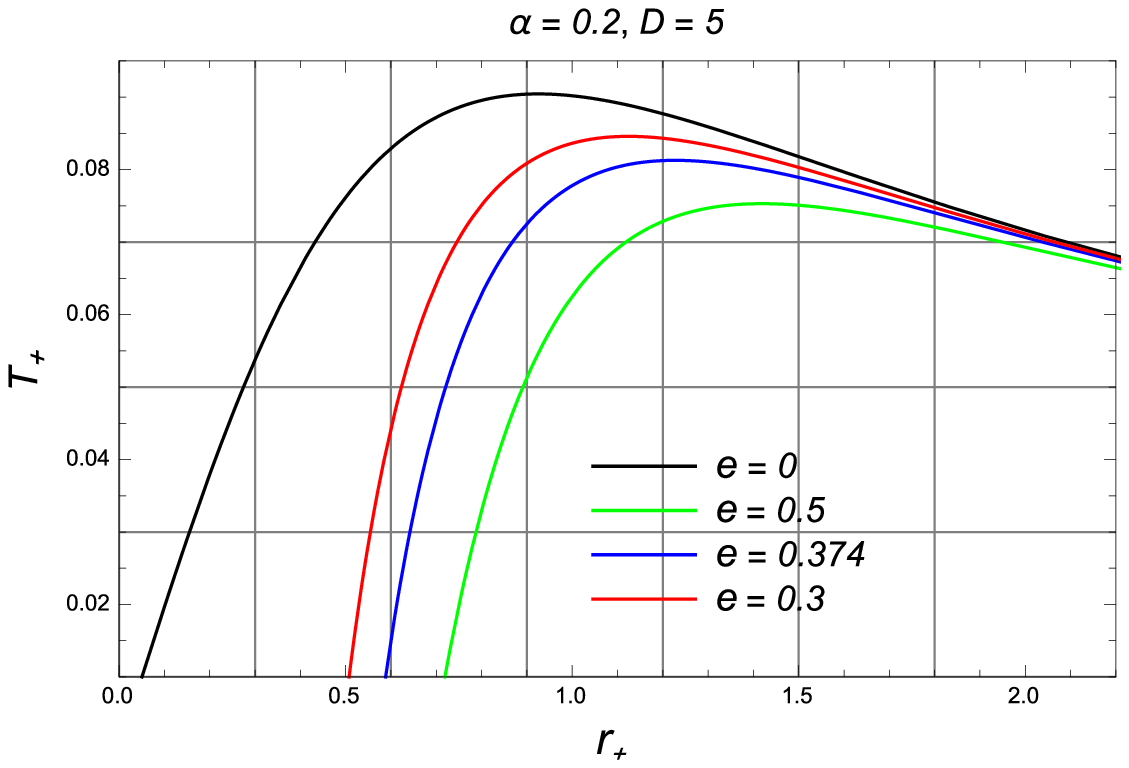}\\
\includegraphics[width=0.5\linewidth]{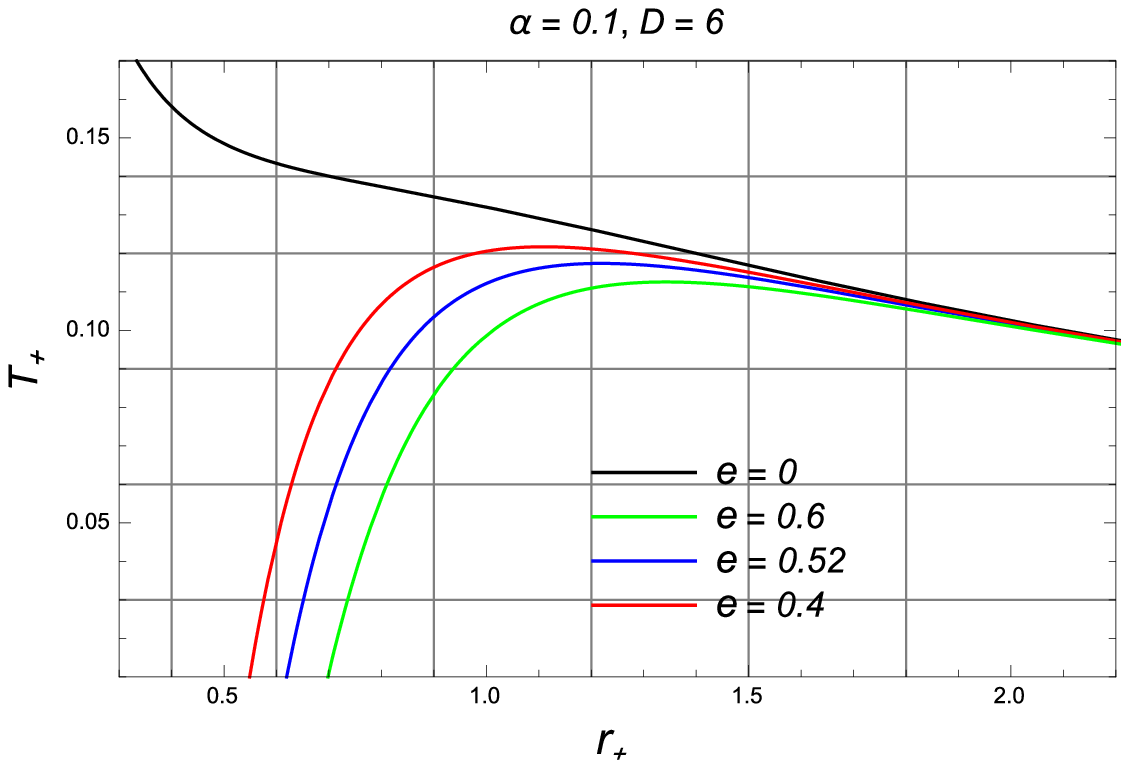}
\includegraphics[width=0.5\linewidth]{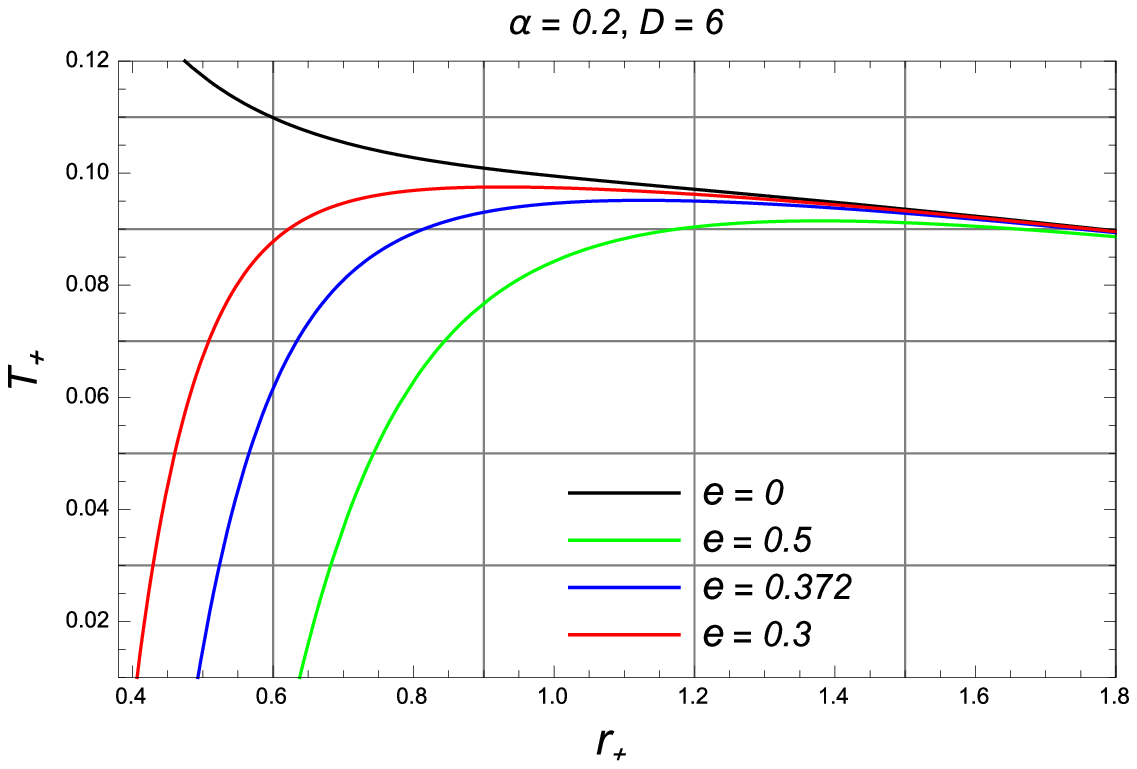}\\
\includegraphics[width=0.5\linewidth]{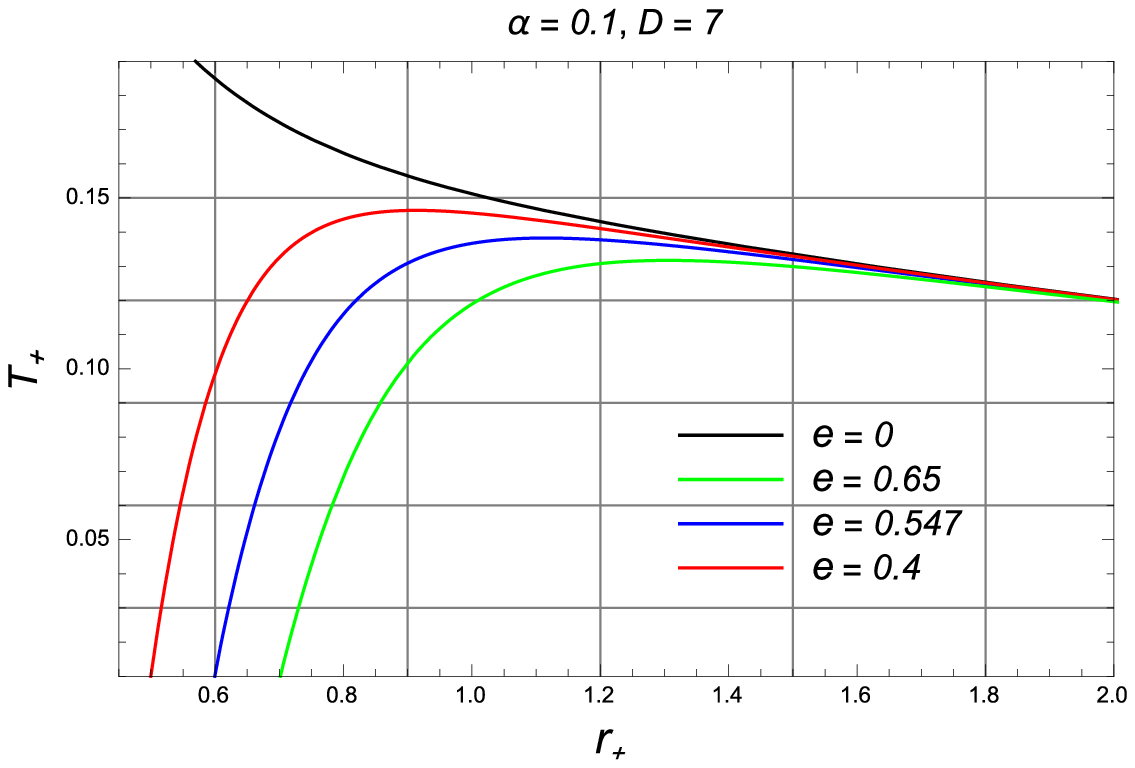}
\includegraphics[width=0.5\linewidth]{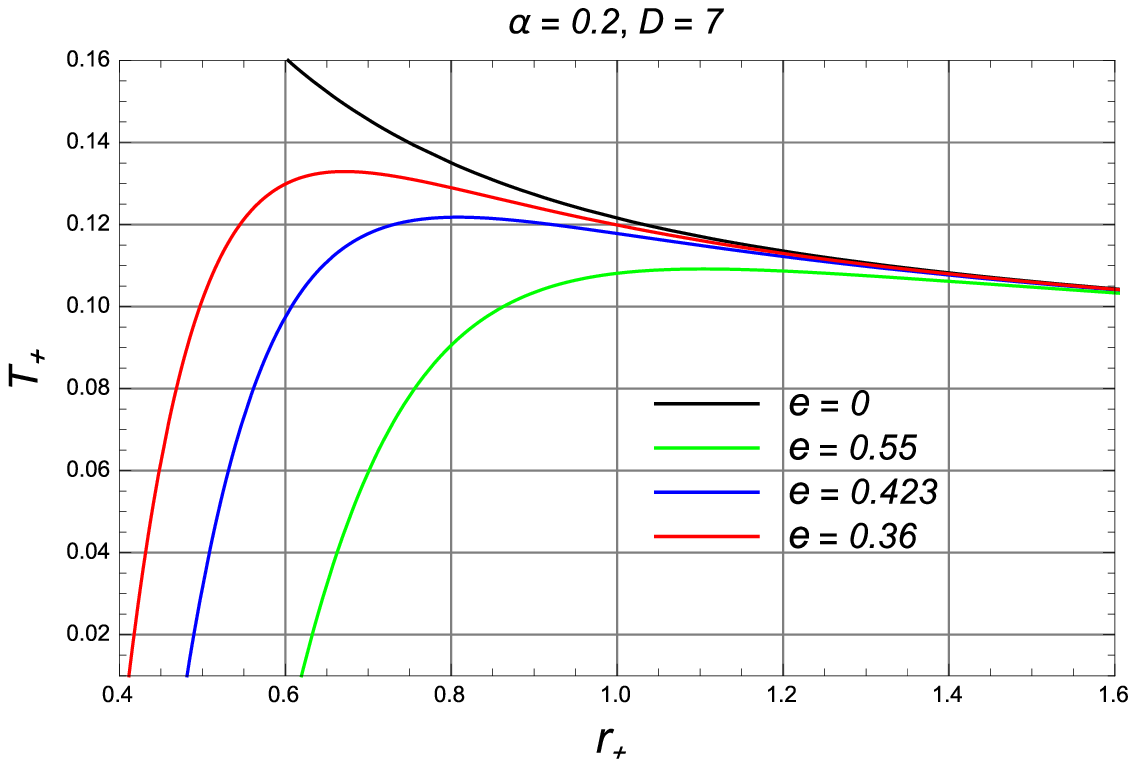}\\
\includegraphics[width=0.5\linewidth]{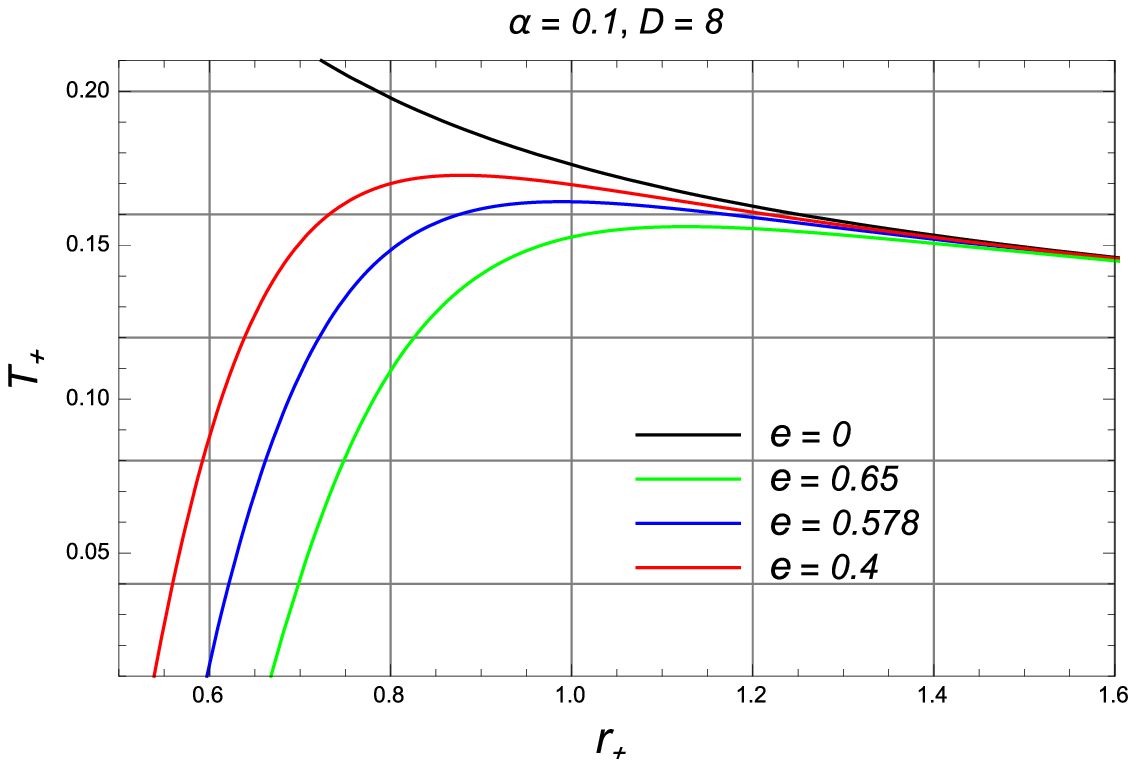}
\includegraphics[width=0.5\linewidth]{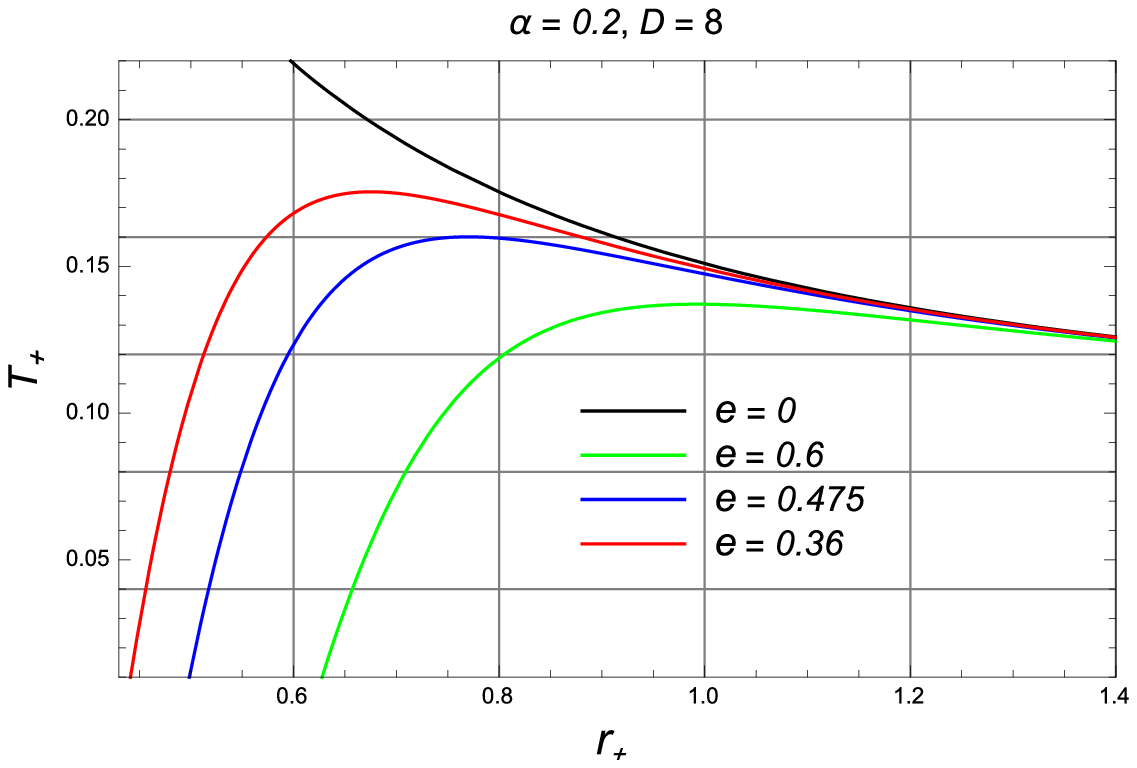}
\end{tabular}
\caption{\label{fig:ht}The Hawking temperature $T_+$ vs horizon radius $r_+$  in various dimensions $D$ = 5, 6, 7, and 8 (top to bottom) for different values of  charge $(e)$ with Gauss-Bonnet coupling parameter $\alpha$ = 0.1 and 0.2 (left to right) with $\mu'=1$. }
\label{rem1}
\end{figure*}
 \begin{figure*} 
\begin{tabular}{c c c c}
\includegraphics[width=0.5\linewidth]{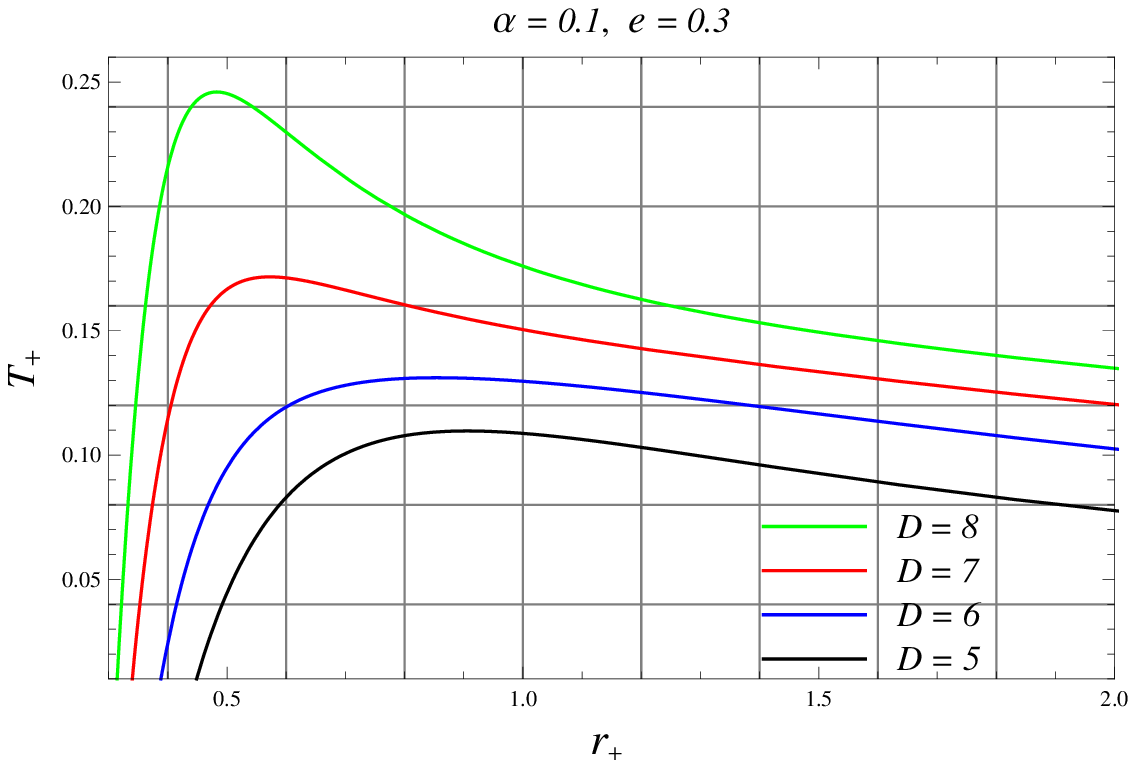}
\includegraphics[width=0.5\linewidth]{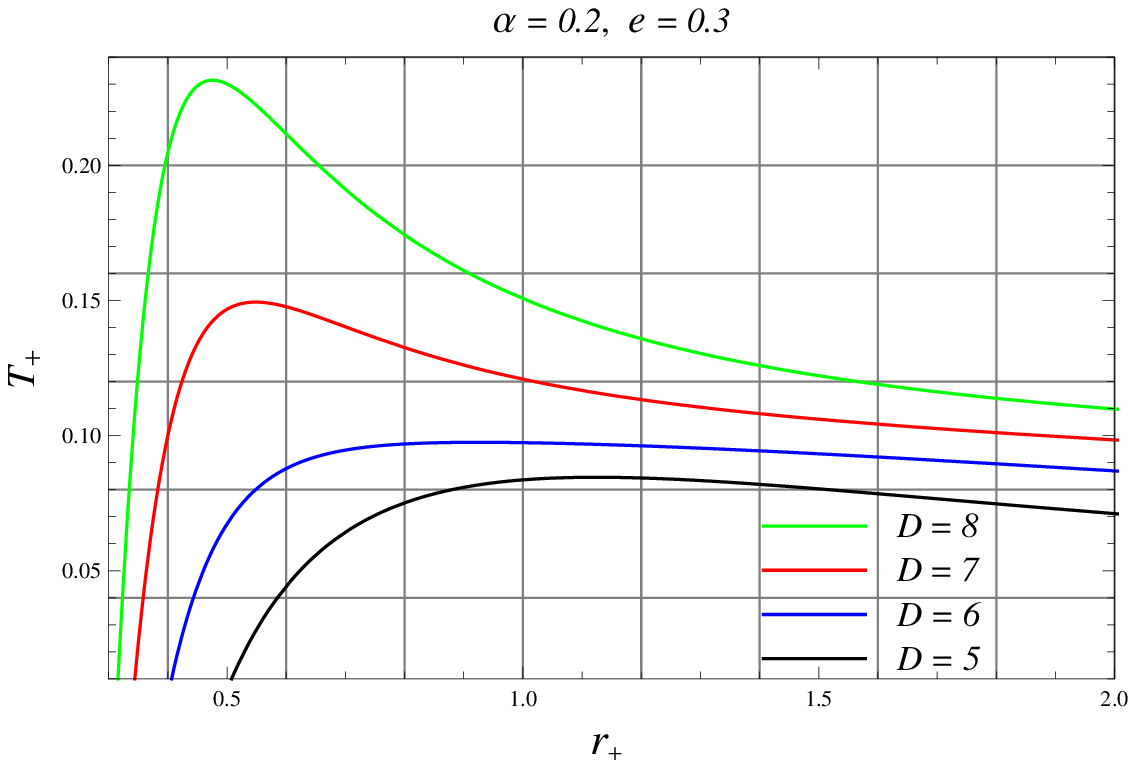}\\
\includegraphics[width=0.5\linewidth]{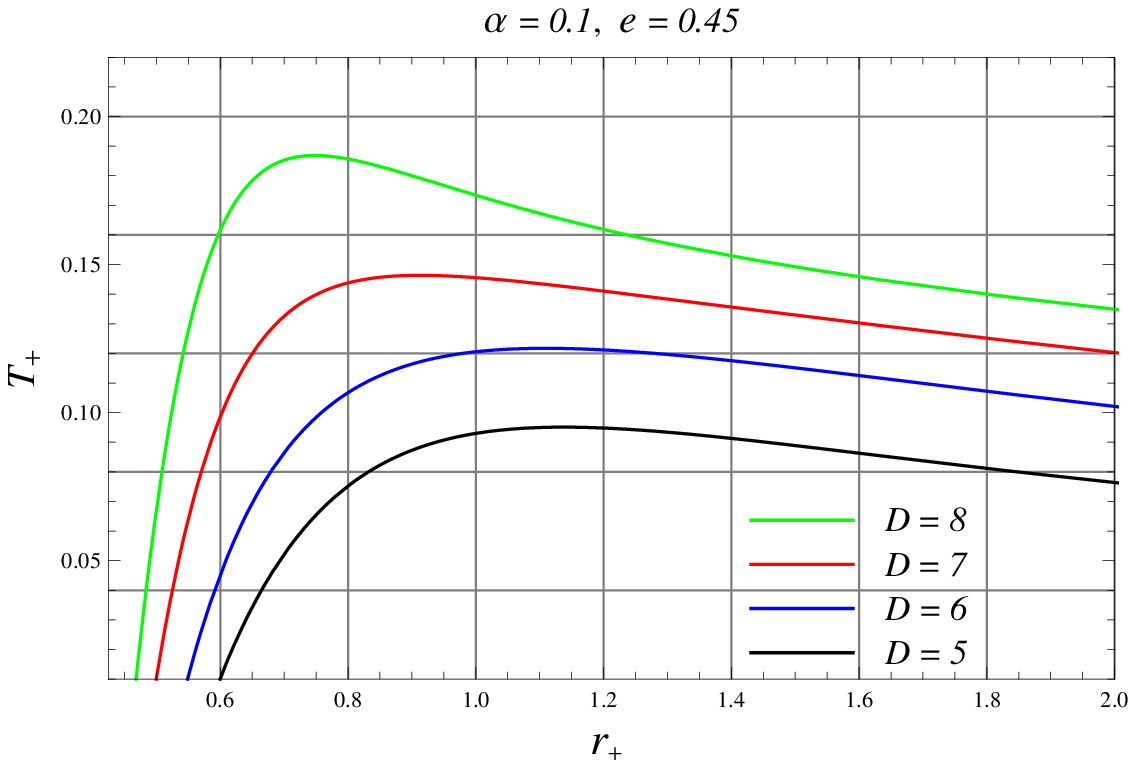}
\includegraphics[width=0.5\linewidth]{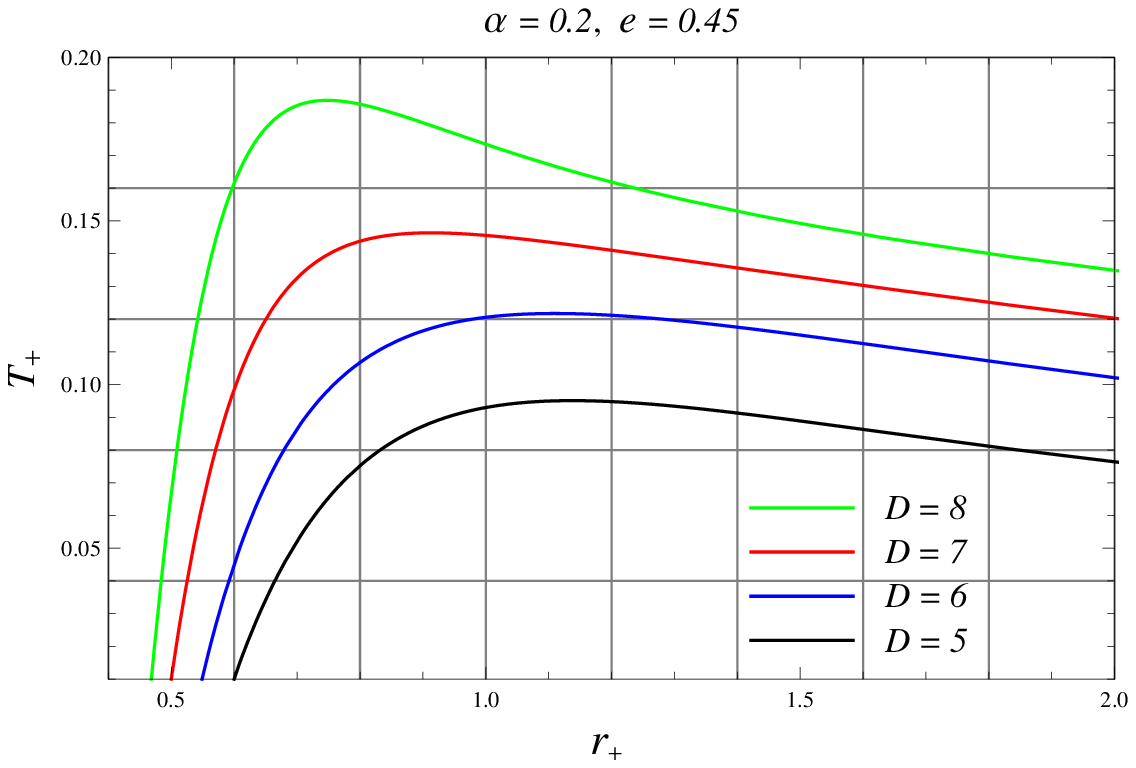}\\
\includegraphics[width=0.5\linewidth]{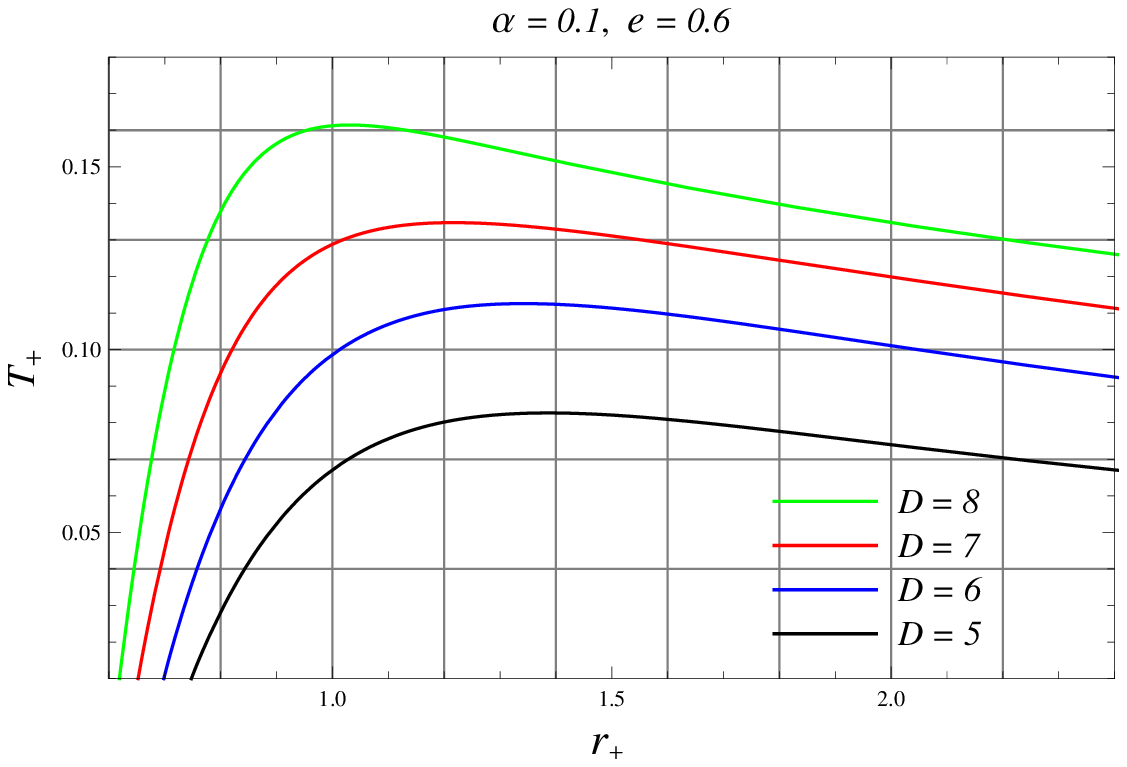}
\includegraphics[width=0.5\linewidth]{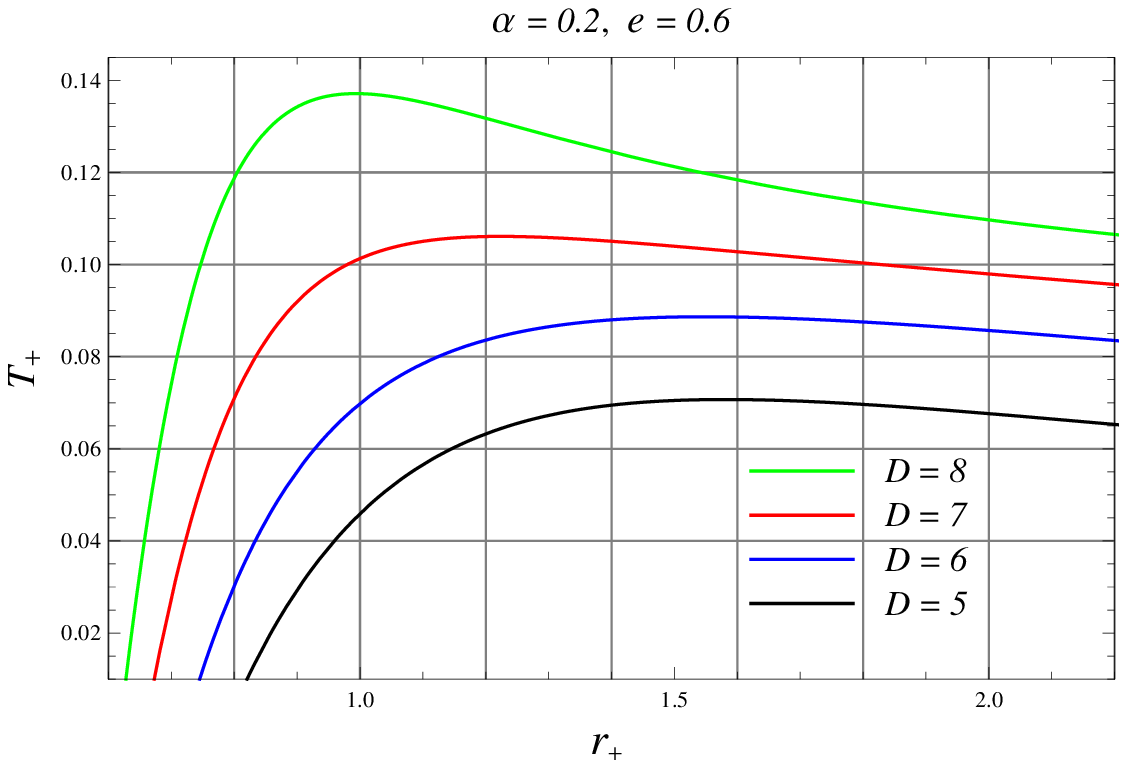}\\
\includegraphics[width=0.5\linewidth]{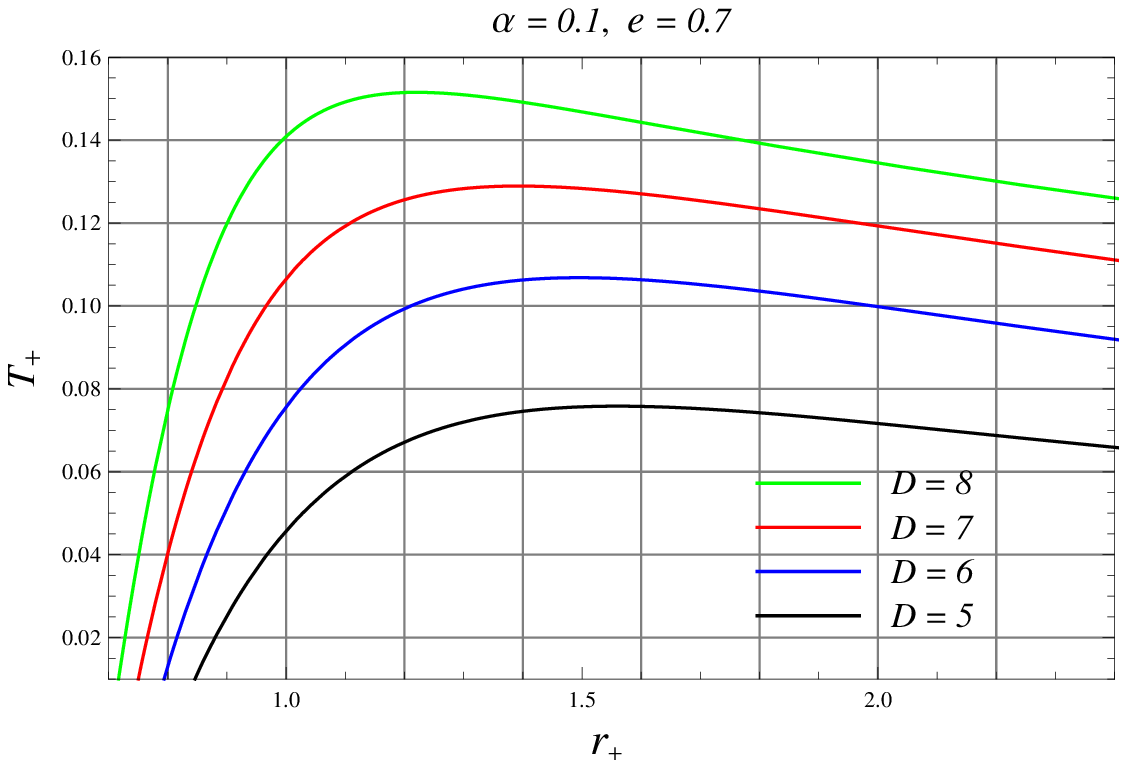}
\includegraphics[width=0.5\linewidth]{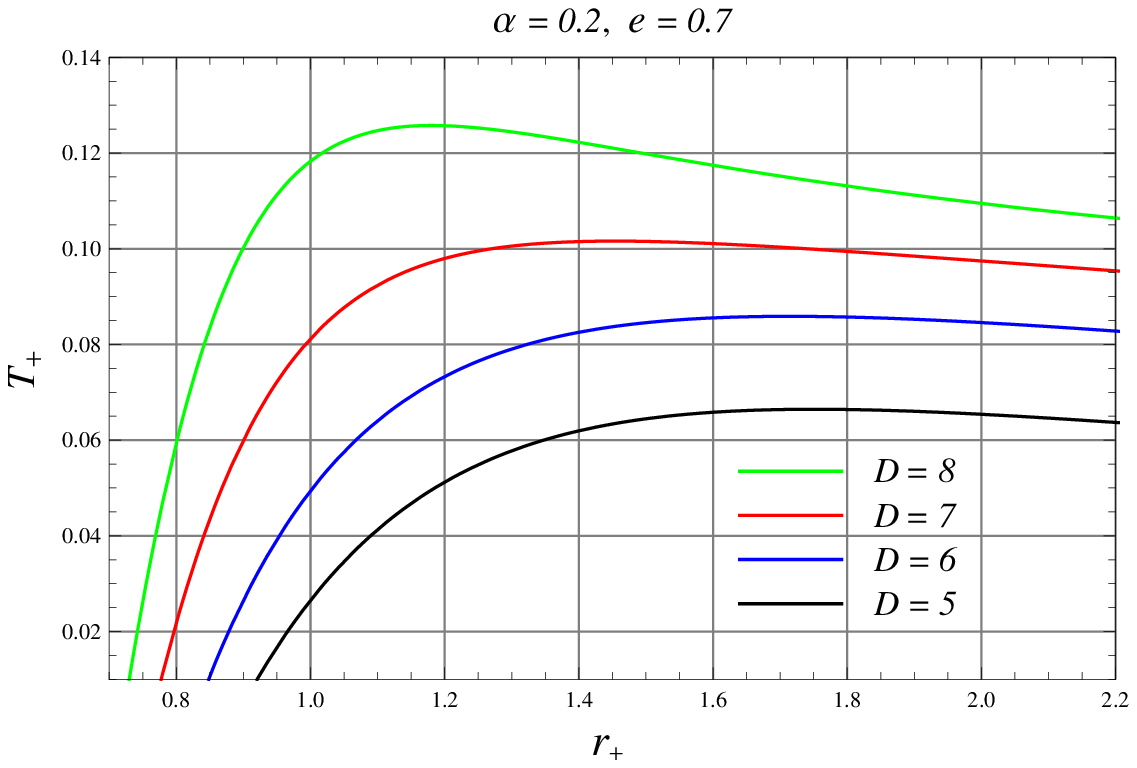}
\end{tabular}
\caption{\label{fig:ht1} The Hawking temperature $T_+$ vs horizon radius $r_+$  for different values of  charge $e$ = 0.3, 0.45, 0.6, 0.7 (top to bottom) with Gauss-Bonnet coupling parameter $\alpha$ = 0.1 and 0.2 (left to right) in various dimensions with $\mu'=1$. }
\label{rem1}
\end{figure*}
By numerical analysis, we conclude that the Hawking temperature vanishes at the radius of the black hole double horizon. The Hawking temperature diverges in the absence of charge ($e=0$), when $r_+ \to 0$ except in 5$D$ (see Fig.  \ref{fig:ht}). However, it becomes finite for non zero value of charge $e$ Fig. \ref{fig:ht}. The temperature depends on both charge $e$ and Gauss-Bonnet parameter $\alpha$.

To calculate an important quantity associated with the black hole, in term of  horizon radius $r_+$, known as entropy, we note that black hole behaves as thermodynamical system; quantities associated must obey first law of thermodynamics \cite{thermo}
\begin{eqnarray}\label{1law}
dM_+ = T_+dS_+ +\Phi de,
\end{eqnarray}
where $S$ is the entropy of the black hole and $\Phi$ is  potential and $e$ is the constant charge. The entropy can be obtained by integrating Eq. (\ref{1law}) as
\begin{eqnarray}\label{sf}
S_+=\int T_+^{-1} dM_+=\int T_+^{-1}\frac{\partial M_+}{\partial r_+}dr_+.
\end{eqnarray}
Now, substituting Eqs. (\ref{M1}) and (\ref{temp1}) in Eq. (\ref{sf}), the entropy of Bardeen-EGB-AdS black hole becomes

\begin{eqnarray}
S_+&=& \frac{(D-2)V_{D-2}(1+\frac{e^{D-2}}{r_+^{D-2}})^\frac{D-1}{D-2}r_+^{D-2}}{4}\Big[(D-4)\frac{r_+^{D-2}}{e^{D-2}}H_1+(D-3)[\frac{2\tilde{\alpha}}{r_+^2}H_2-(D-4)H_3]\nonumber\\&&-(D-3)(D-4)\tilde{\alpha} \frac{e^{D-2}}{r_+^D}H_4\Big],
\label{entropy1}
\end{eqnarray}
 with
\begin{eqnarray}
 &&H_1 =  _{2}F_1\left[1, 2,\frac{-(2D-5)}{(D-2)}, -\frac{r_+^{D-2}}{e^{D-2}}\right],\qquad\
H_2 = _{2}F_1\left[1,\frac{3}{(D-2)},\frac{2}{(D-2)}, -\frac{e^{D-2}}{r_+^{D-2}}\right],\nonumber\\ 
&& H_3 = _{2}F_1\left[1, 1,\frac{(D-3)}{(D-2)},-\frac{r_+^{D-2}}{e^{D-2}}\right],\quad\text{and}\quad
 H_4 = _{2}F_1\left[1, \frac{(D+1)}{(D-2)},\frac{ D}{(D-2)},-\frac{e^{D-2}}{r_+^{D-2}}\right].\nonumber\\
\end{eqnarray}
Where $_{2}F_1$ is the hyper geometric function. The entropy (\ref{entropy1}) reduces to the entropy of EGB-AdS black hole \cite{Neu,cai,Ads,MS1} in the absence of charge ,
\begin{equation}
S_{+} = \frac{(D-2)V_{D-2}r_+^{D-4}}{4}\left[\frac{r_+^2}{D-2}+\frac{2\tilde{\alpha}}{D-4}\right].
\end{equation}
 We recover the entropy of the EGB black hole \cite{sus} when $e=0, \text{and}\, \Lambda=0$  and further we obtained the entropy for $D$-dimensional Bardeen black hole \cite{sabir} in the limit $\alpha \to 0$ 

\begin{eqnarray}
S_+&=& \frac{(D-2)V_{D-2}(1+\frac{e^{D-2}}{r_+^{D-2}})^\frac{D-1}{D-2}r_+^{D-2}}{4}\left[(D-4)\frac{r_+^{D-2}}{e^{D-2}}H_1-(D-3)(D-4)H_3\right],
\label{entropy3}
\end{eqnarray}
The Eq. (\ref{entropy1}) reduce to the entropy of  Schwarzschild-Tangherlini black hole \cite{sus,PK} when $e=0$, $\alpha \to 0$.  The entropy for our model differs from the expression for entropy in general relativity, in which it is proportional to the area of the event horizon \citep{jdb}. However, it is interesting to note that the expression for entropy of the  black hole is independent of cosmological constant.

\section{Local Stability and Black Hole Remnants}
In order to analyze local stability, we shall consider the specific heat of Bardeen-EGB-AdS black holes. The heat capacity of the black hole is given \cite{cai,sus}
\begin{eqnarray}\label{SH}
C_+&=&\frac{\partial{M_+}}{\partial{T_+}}=\left(\frac{\partial{M_+}}{\partial{r_+}}\right)\left(\frac{\partial{r_+}}{\partial{T_+}}\right).
\end{eqnarray}

The region of the parameter space where the specific heat is positive, the black hole are locally stable \cite{sus} to thermal fluctuations. Thus, when the specific heat is positive, then increase in the black hole temperature will result an increase in the entropy thereby giving the thermodynamic stable configuration. This is because a black hole at higher temperature is stable, while unstable at low temperature \cite{sus}.  The heat capacity of Bardeen-EGB-AdS black hole reads
\begin{eqnarray}\label{SH1}
C_+ &=&\frac{(D-2)  V_{D-2}(1+\frac{e^{D-2}}{r^{D-2}_+})^\frac{2D-3}{D-2}(r^2_++2\tilde{\alpha})^2r^{D-4}_+}{4\left(A(\frac{e^{D-2}}{r^{D-2}_+})^2+B\frac{e^{D-2}}{r^{D-2}_+}-C+E\frac{1}{l^2}\right)}\nonumber
\\&&\qquad\qquad\qquad\qquad\left[(D-3)r^2_++(D-5)\tilde{\alpha}-2\frac{e^{D-2}}{r^{D-2}_+}(r^2_++2\tilde{\alpha})+\frac{D-1}{l^2}r_{+}^4\right], 
\end{eqnarray}
where
\begin{eqnarray}
&&A=2(r^{2}_++2\tilde{\alpha})^2,\nonumber\\
&&B = (D^2-4D+7)r^4_++(3D^2-10D+23)\tilde{\alpha}r^2_++2(D^2-4D+11)\tilde{\alpha}^2,\nonumber\\
&&C = (D-5)(3r^2_++2\tilde{\alpha})\tilde{\alpha}+(D-3)(r^2_+-2\tilde{\alpha})r^2_+,\nonumber\\
&&E = \left[(D-1)^2r^6_++2(D^2-1)\tilde{\alpha}r^4_+\right]\frac{e^{D-2}}{r^{D-2}_+}+(D-1)(r^2_++6\tilde{\alpha})r^4_+.
\end{eqnarray}
It can be seen clearly, that the heat capacity depends on the Gauss-Bonnet coefficient ${\alpha}$,  charge $e$, cosmological constant $\Lambda$ and the dimension $D$. In the absence of  charge $e$, we reduce to the expression for heat capacity of EGB-AdS black hole \cite{Neu,cai,Ads,MS1}, which reads 
\begin{eqnarray}
C_{+}&=& \frac{(D-2)  V_{D-2}(r^2_++2\tilde{\alpha})^2\left[(D-3)r^2_++(D-5)\tilde{\alpha}+\frac{D-1}{l^2}r_{+}^4\right]r^{D-4}_+ }{4\left[(D-3)(r^2_+-2\tilde{\alpha})r^2_+-(D-5)(3r^2_++2\tilde{\alpha})\tilde{\alpha}+\frac{(D-1)(r^2_++6\tilde{\alpha})}{l^2}r^4_+\right]},
\end{eqnarray}
which is also the specific heat of EGB black hole when switch off the charge $e=0$ \cite{sus,Wij}.
We recover the $D$-dimensional Bardeen black hole \cite{sabir} specific heat in the limit $\alpha=\Lambda= 0$  
\begin{eqnarray}
C_{+}&=& \frac{(D-2)  V_{D-2}(1+\frac{e^{D-2}}{r^{D-2}_+})^\frac{2D-3}{D-2}\left[(D-3)-\frac{2e^{D-2}}{r^{D-2}_+}\right]r^{D-2}_+ }{4\left[2(\frac{e^{D-2}}{r^{D-2}_+})^2+(D^2-4D+7)\frac{e^{D-2}}{r^{D-2}_+}-(D-3)\right]}.
\end{eqnarray}
 We can find the specific heat for Schwarzschild Tangharhelini black hole in the limit $e=0$ and $\Lambda= \alpha=0$ \cite{sus} 
\begin{equation}
C_{+} =-\frac{{(D-2)V_{D-2}}r^{D-2}_+}{4}.
\end{equation}

\begin{figure*} 
\begin{tabular}{c c c c}
\includegraphics[width=0.5\linewidth]{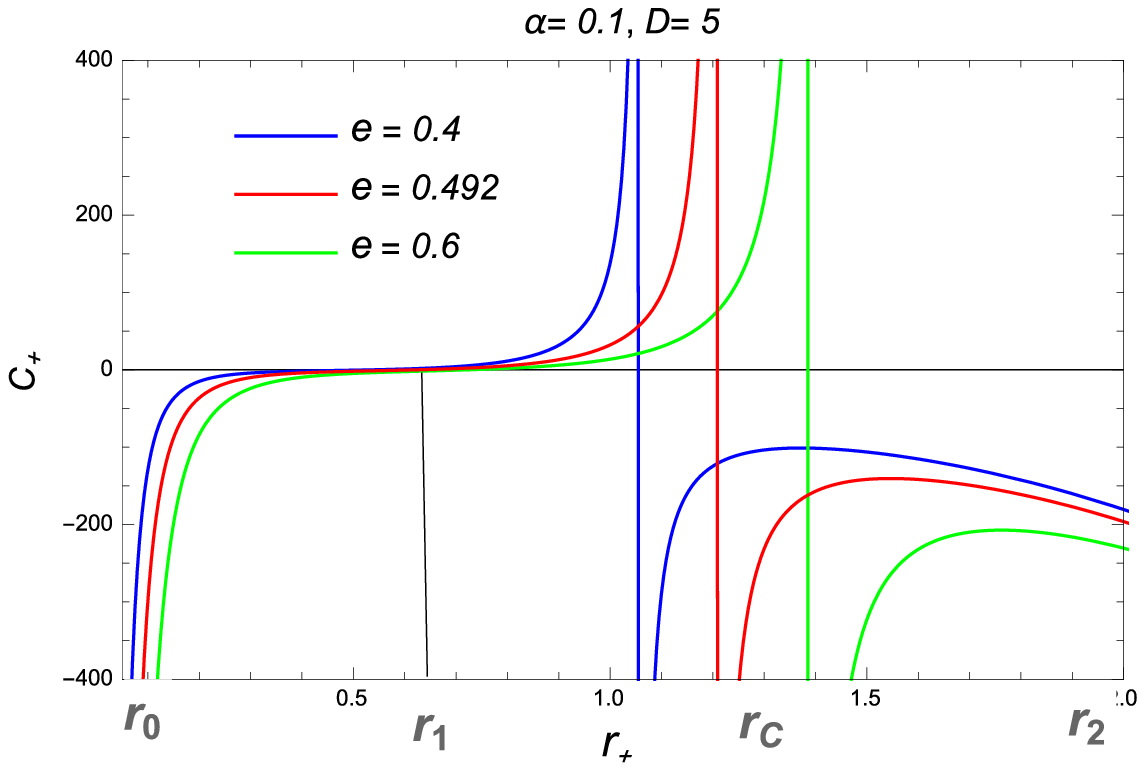}
\includegraphics[width=0.5\linewidth]{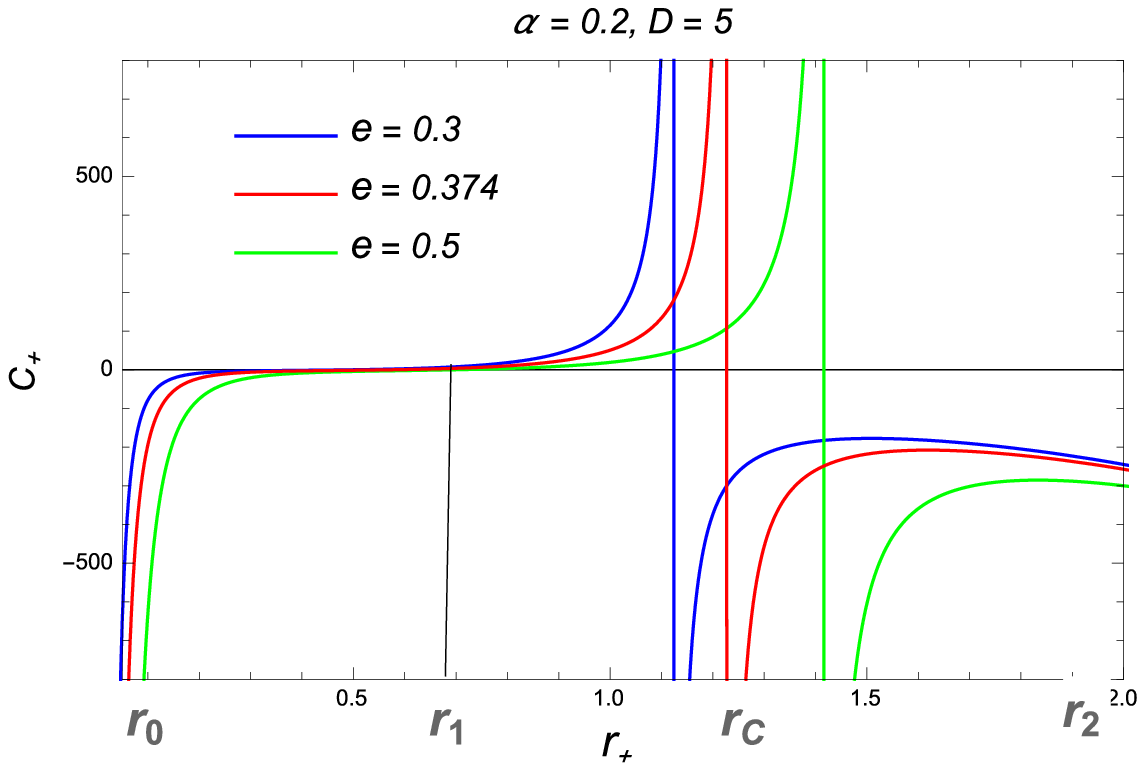}\\
\includegraphics[width=0.5\linewidth]{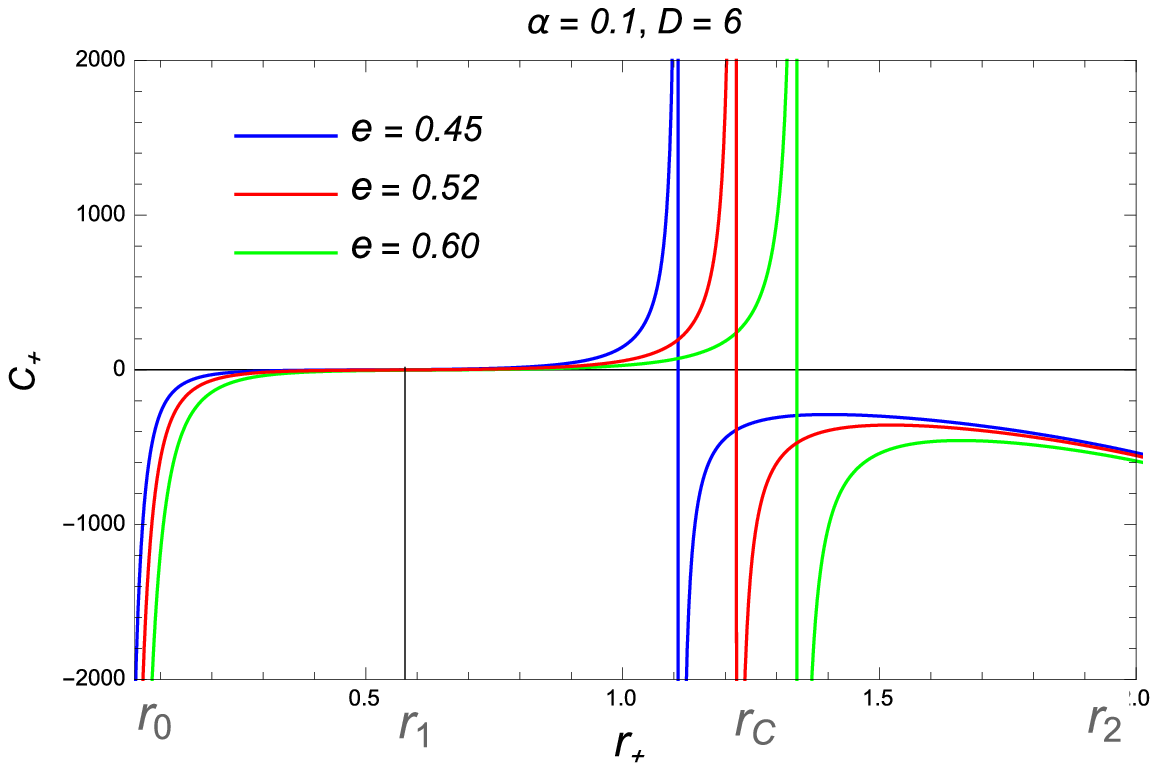}
\includegraphics[width=0.5\linewidth]{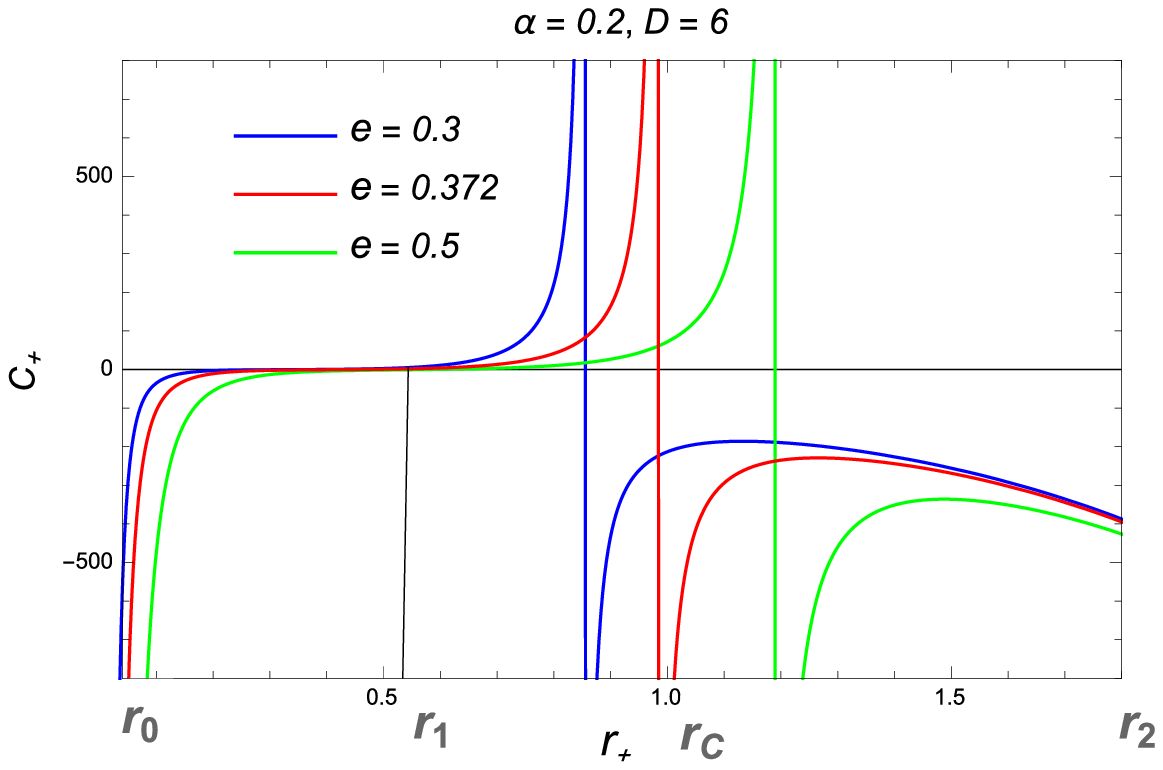}\\
\includegraphics[width=0.5\linewidth]{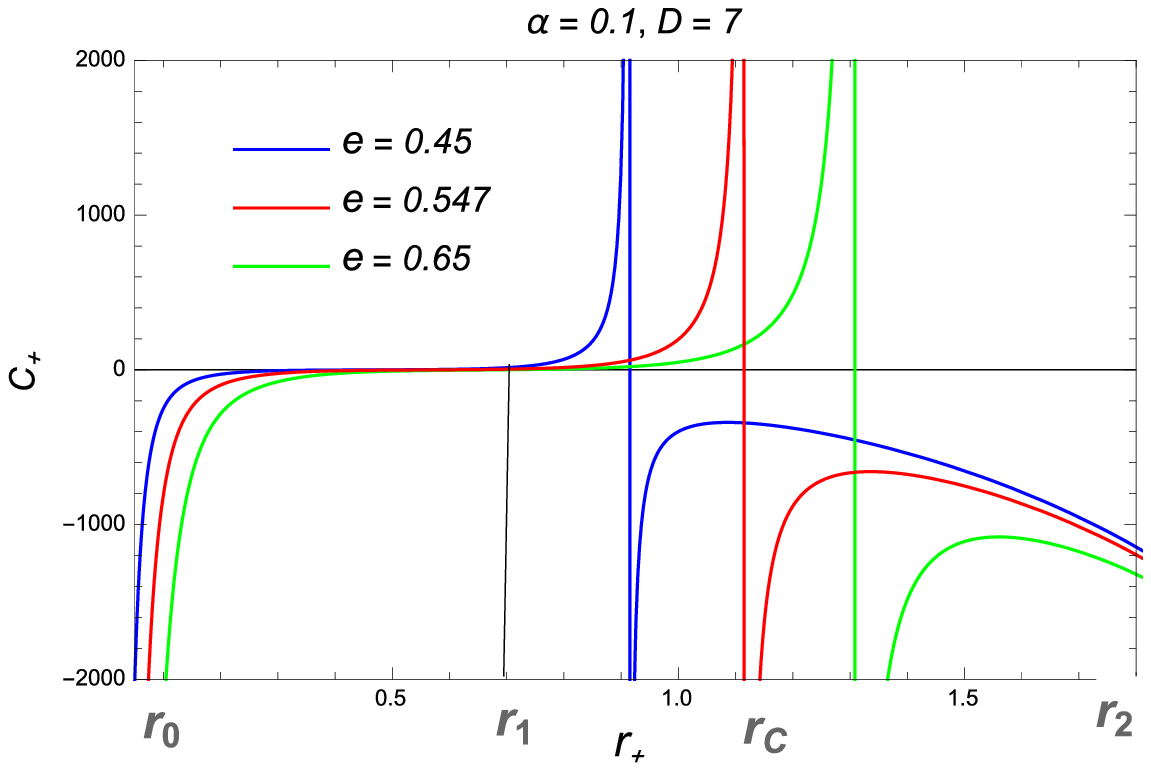}
\includegraphics[width=0.5\linewidth]{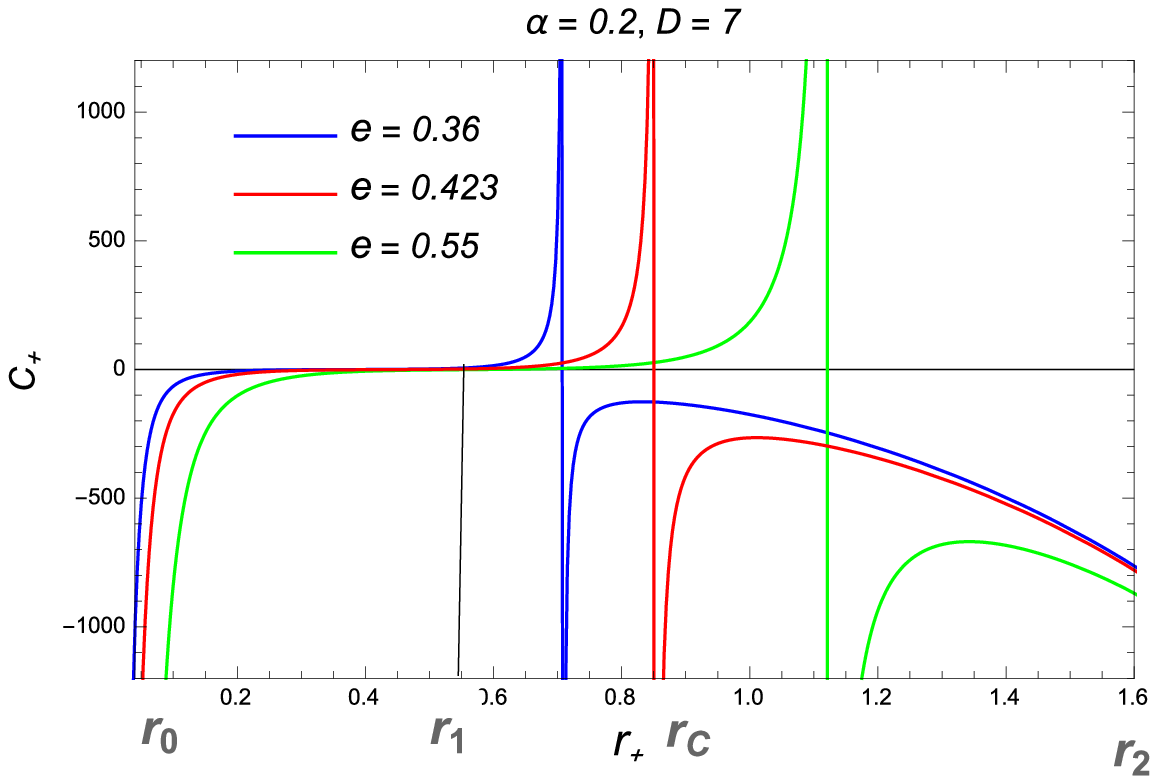}\\
\includegraphics[width=0.5\linewidth]{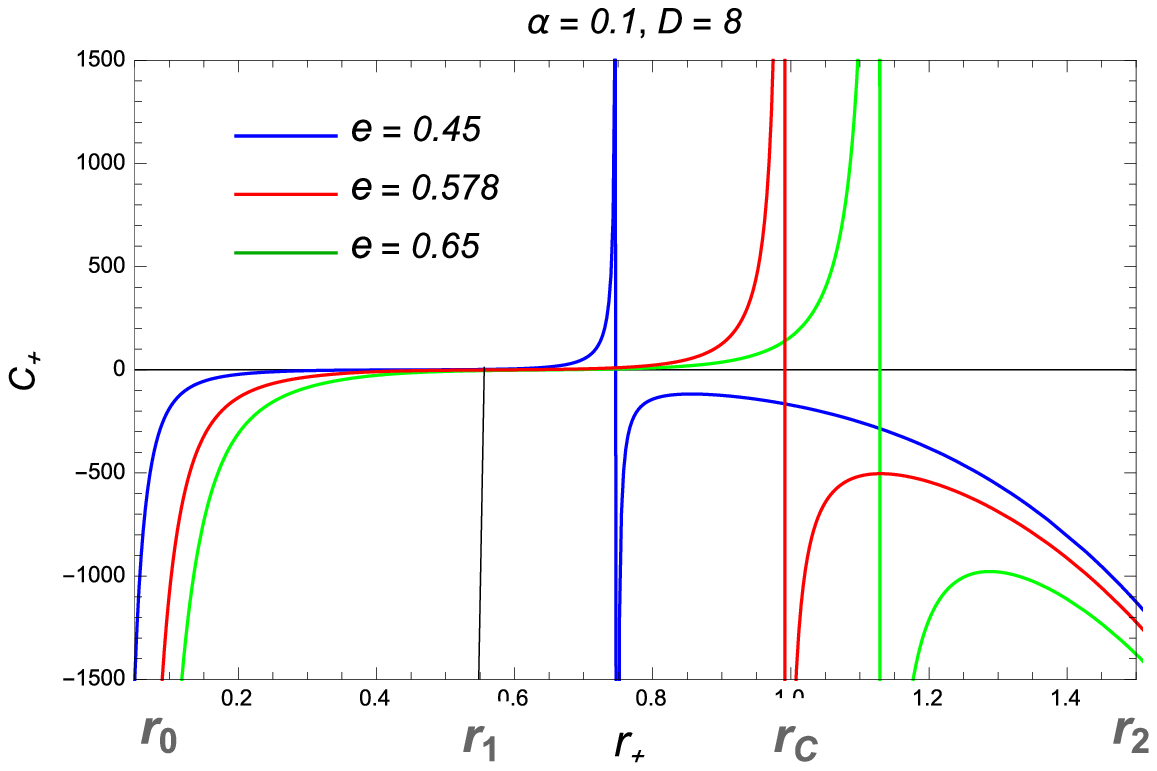}
\includegraphics[width=0.5\linewidth]{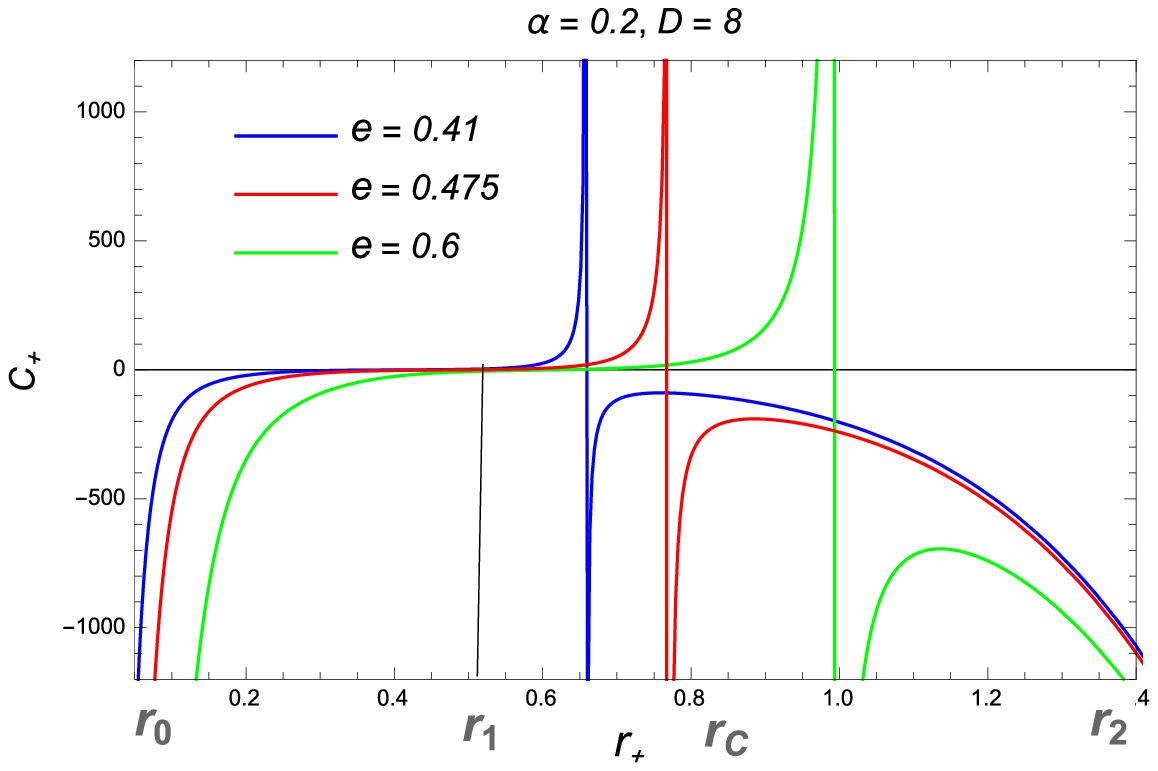}
\end{tabular}
\caption{\label{fig:st}The specific heat $C_+$ vs horizon radius $r_+$ in various dimensions $D$ = 5, 6, 7, and 8 (top to bottom) for different values of charge $e$ with Gauss-Bonnet coupling parameter $\alpha$ = 0.1 and 0.2 (left to right) with $\mu'=1$. }
\end{figure*}
 In what follows, we analyze the stability of the EGB AdS black hole and bring out the effect of the nonlinear electrodynamics background. Due to the complexity of  Eq.~(\ref{SH1}), it is difficult to analyze the heat capacity analytically; hence, we plot it in Fig. \ref{fig:st} for different values of parameters in different dimensions.
Clearly, the positivity of heat capacity $C_+>0$ of the black hole is sufficient to state that the black hole is thermodynamically stable. Fig. \ref{fig:st} shows that heat capacity is discontinuous exactly at one point for a given value of $e$ and $\alpha$, which is identified as the critical radius $r^{C}_+$. Further, we noticed that the heat capacity changes its sign around $r^{C}_+$. Thus, we can say the black hole is thermodynamically stable for  $r_1<r_+ < r^{C}_+$, whereas it is thermodynamically unstable for $r_+>r_+^C$, and there is a second order phase transition at $r_+=r^{C}_+$ from the stable to unstable phases. So, the heat capacity of Bardeen-EGB-AdS black hole, in any dimension for different values of $e$, $\alpha$ and $\Lambda$, is positive for $r_1<r_+ < r^{C}_+$, and it is negative for $r_1>r_+>r_+^C$. Here, we noticed from the Fig. \ref{fig:st} that the value of critical radius $r^{C}_+$ increases with the increase in the charge $e$, for given value of Gauss-Bonnet coupling constant $\alpha$ and cosmological constant. Thus, the change in the value of charge affected the thermodynamical stability of the black hole. One can find that the Bardeen-EGB-AdS black holes have two unstable regions and a stable region as shown in Table \ref{stable}

\begin{table}[h]
    \begin{center}        
        \begin{tabular}{|l| l l l   |  l l l  |  }
            \hline
        
           \,\, \,\,Region &$$ & ~~~State  &~~~$$  &~~ $$ &~~~Stability  &~~~~ $$   \\
            \hline

            \,\,\,\,$r_0<r_+<r_1$&\,\, & ~~~small  &   \,\,   &  &~~~unstable&      \\
\,\,\,\,$r_1<r_+<r_C$&\,\, &~~~intermediate  &     \,\,   & &~~~stable&     \\
            \,\,\,\,$r_+>r_C$& \,\,&~~~large  &  \,\,  &  &~~~unstable& \,\,         \\
            \hline 
        \end{tabular}
       \end{center}
       \caption{The state and stability of black hole with horizon radius $r+$}
       \label{stable}
\end{table}

The global stability of  the black hole can be  the study by the behaviour of free energy. The Gibb's free energy of black hole can be defined as \cite{Herscovich}
\begin{eqnarray}\label{fe}
G_+&=&M_+-T_+S_+,
\end{eqnarray} 
substituting Eqs. (\ref{M1}) and (\ref{temp1}) into Eq. (\ref{fe}), we get the expression for Gibb's free energy of Bardeen-EGB-AdS black hole, which reads
\begin{eqnarray}\label{fe1}
G_+&=&\frac{(D-2) V_{D-2}r_+^{D-3}(1+\frac{e^{D-2}}{r^{D-2}})^\frac{D-1}{D-2}}{16 \pi}\Big[(1 + \frac{\tilde{\alpha}}{r_{+}^2}+\frac{r_{+}^2}{l^2})\nonumber\\
&&-\frac{(D-3)r_{+}^2+(D-5)\tilde{\alpha}-2\frac{e^{D-2}}{r_{+}^{D-2}}(r_{+}^2+2\tilde{\alpha}+\frac{D-1}{l^2}r_{+}^4)}{ (r_{+}^2+2\tilde{\alpha})(1+\frac{e^{D-2}}{r_{+}^{D-2}})}\Big[(D-4)\frac{r_+^{D-2}}{e^{D-2}}H_1+(D-3)\nonumber\\
&&\left(\frac{2\tilde{\alpha}}{r_+^2}H_2-(D-4)H_3\right)-(D-3)(D-4)\tilde{\alpha}\frac{e^{D-2}}{r_+^D}H_4\Big]\Big].
\end{eqnarray}
\begin{figure*} 
\begin{tabular}{c c c c}
\includegraphics[width=0.5\linewidth]{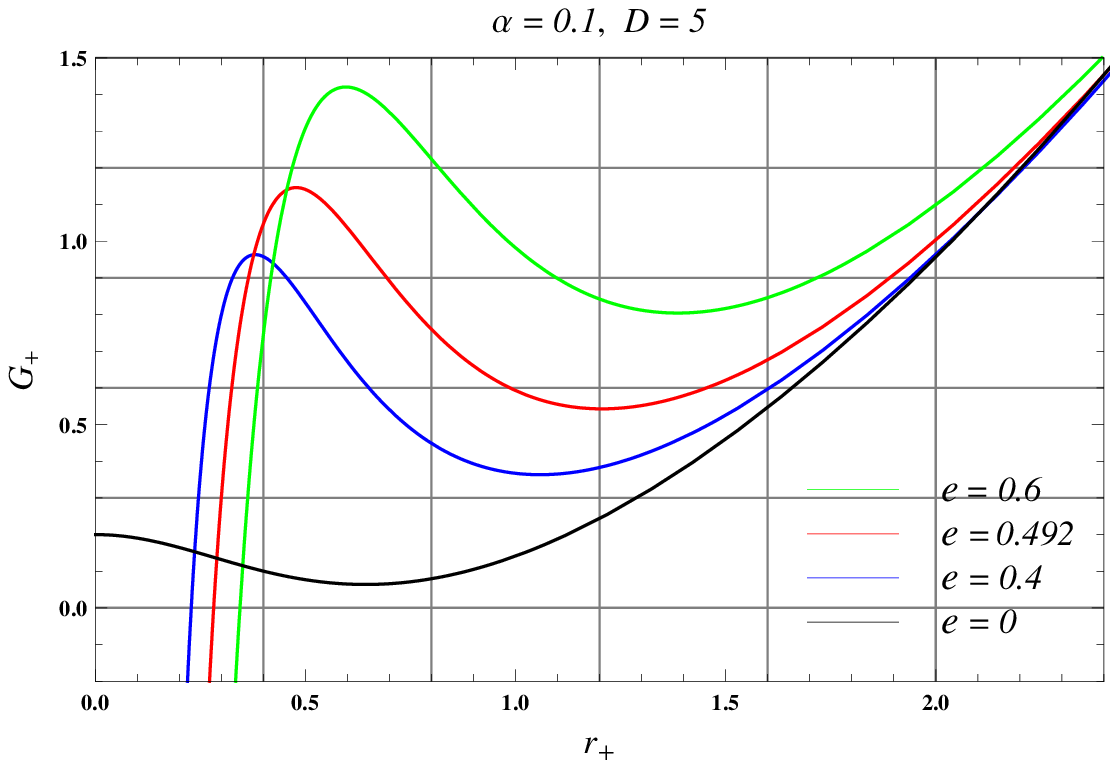}
\includegraphics[width=0.5\linewidth]{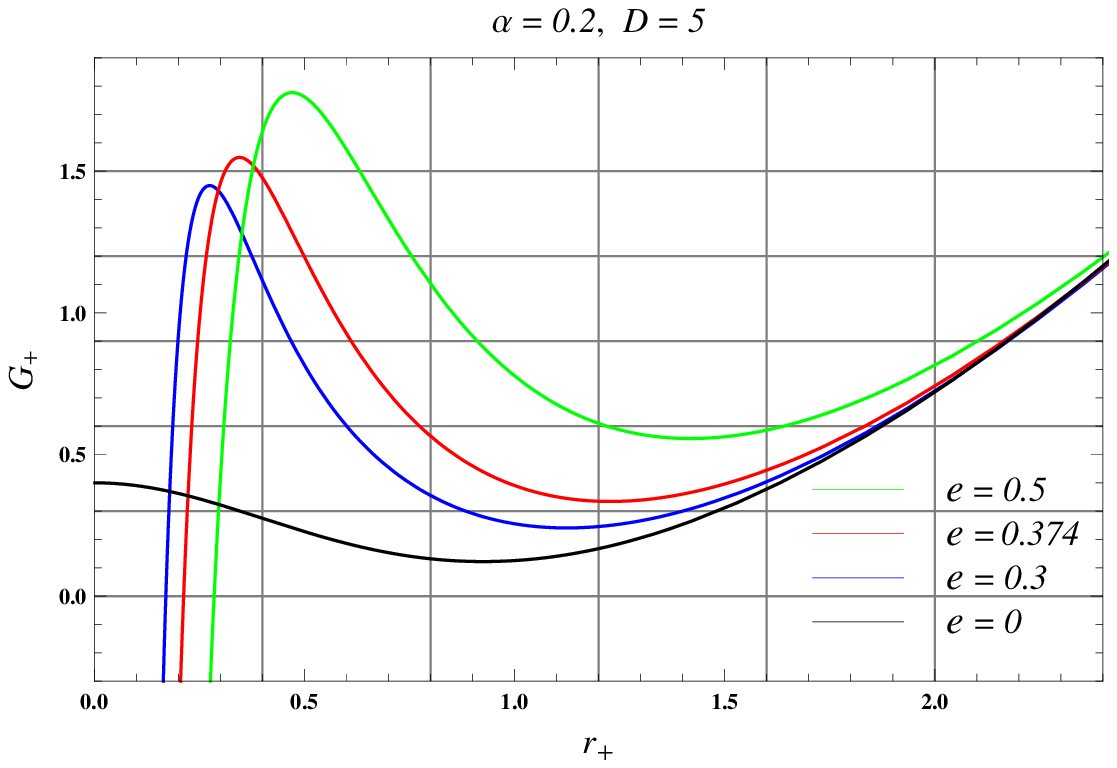}\\
\includegraphics[width=0.5\linewidth]{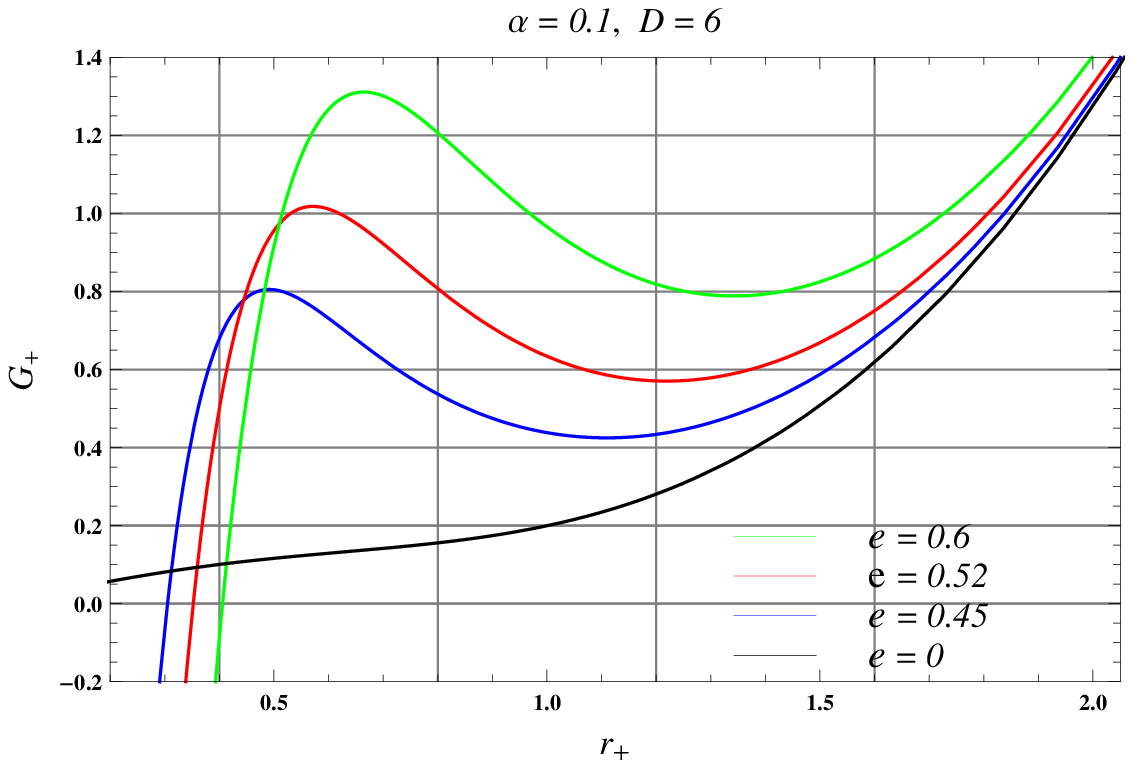}
\includegraphics[width=0.5\linewidth]{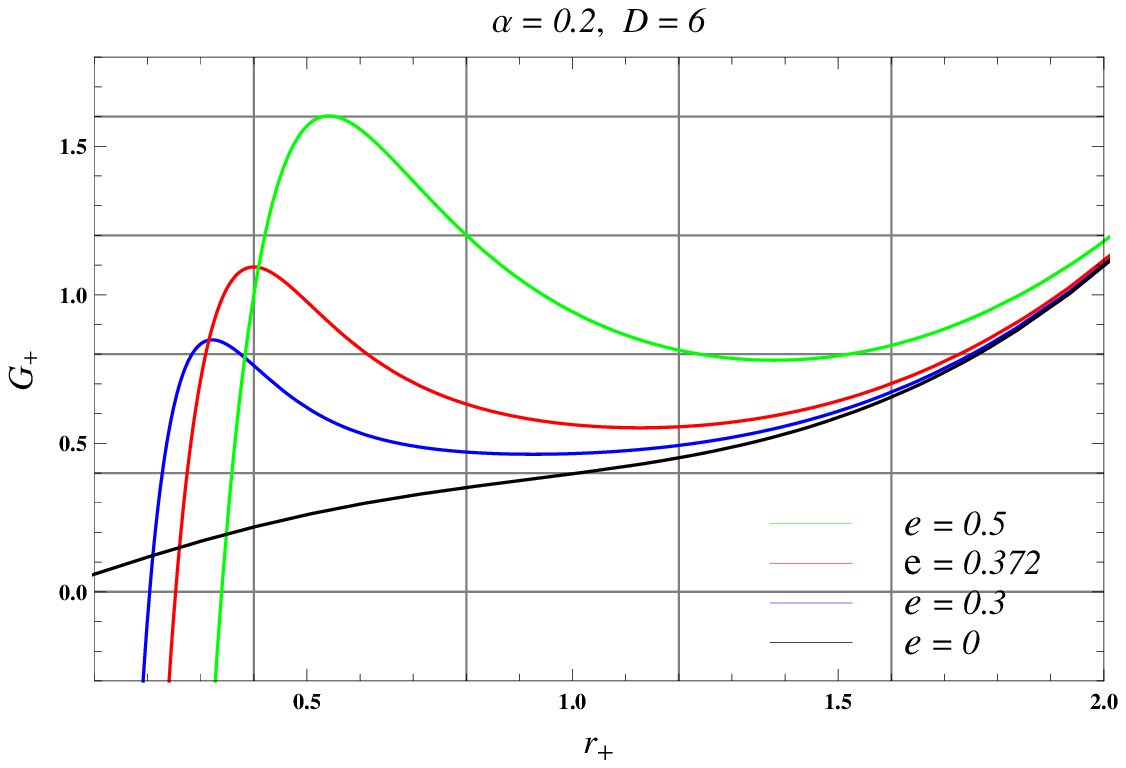}\\
\includegraphics[width=0.5\linewidth]{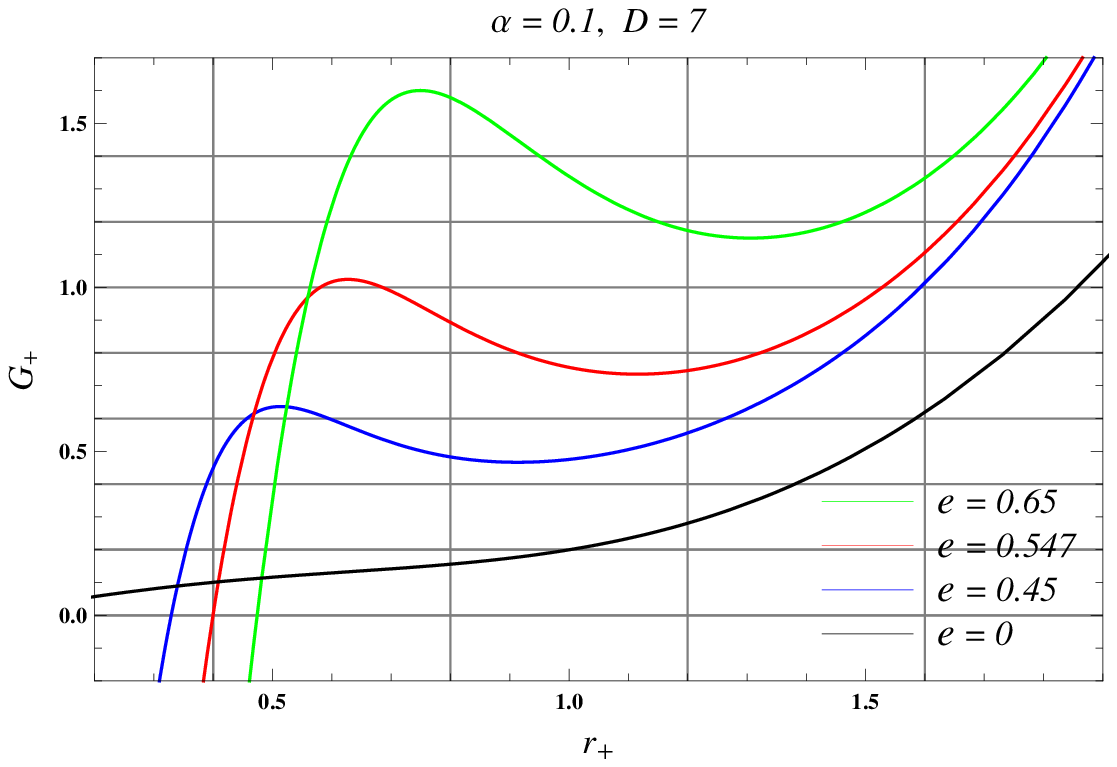}
\includegraphics[width=0.5\linewidth]{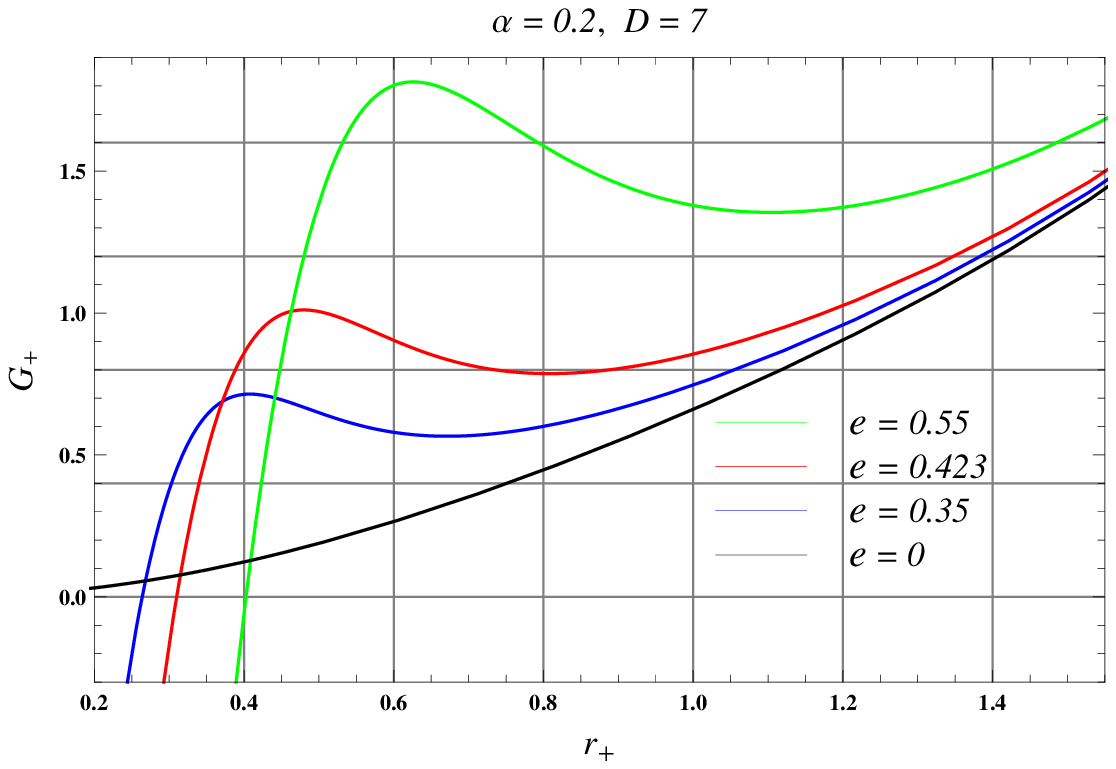}\\
\includegraphics[width=0.5\linewidth]{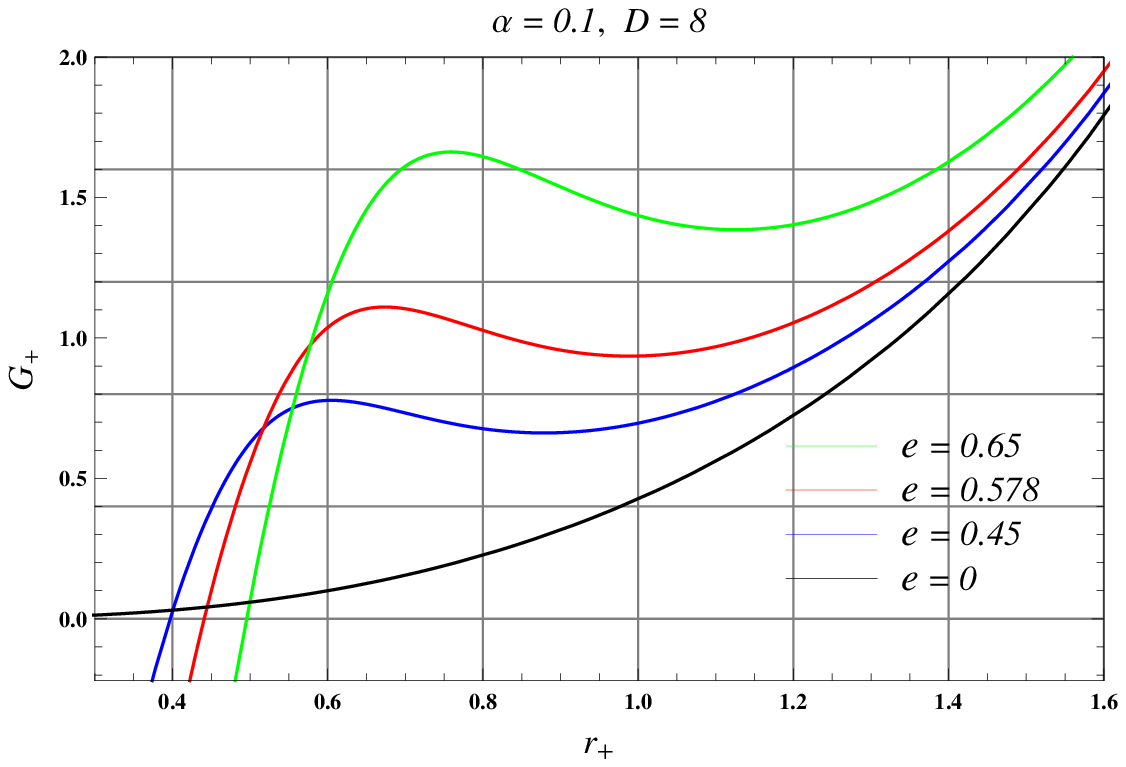}
\includegraphics[width=0.5\linewidth]{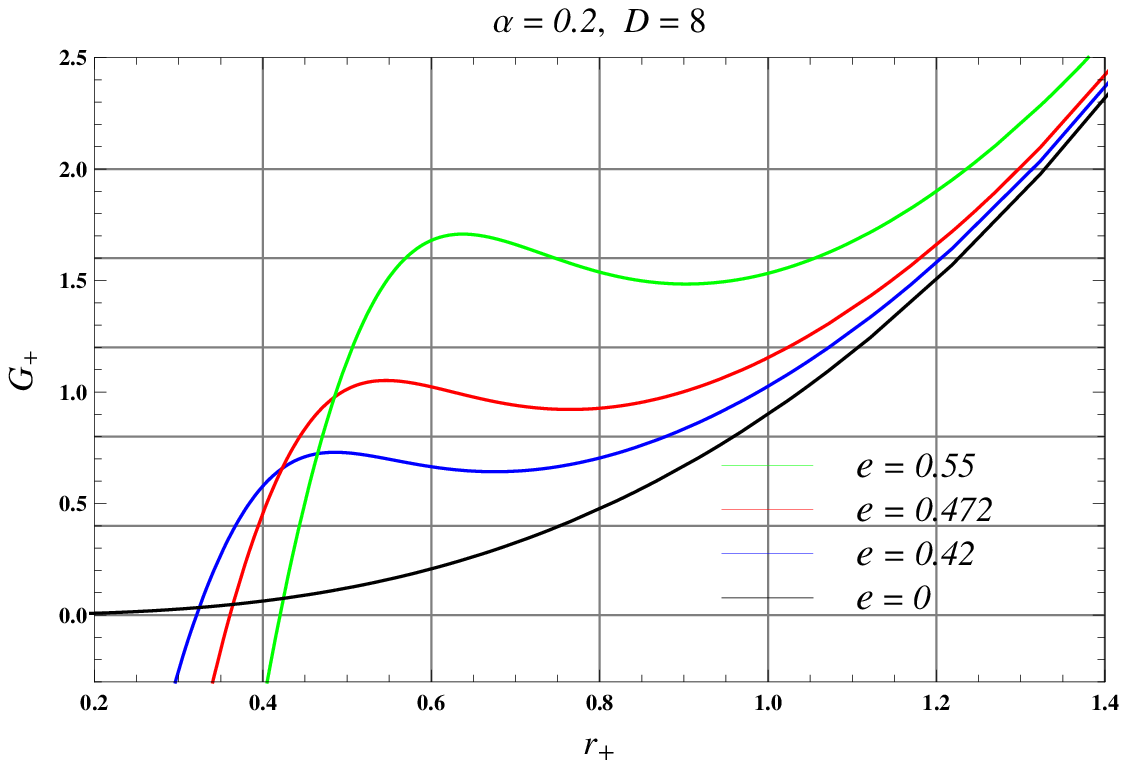}
\end{tabular}
\caption{\label{fig:fe} Gibb's free energy $G_+$ vs horizon radius $r_+$ in various dimensions $D$ = 5, 6, 7, and 8 (top to bottom) for different values of charge ($e$) with Gauss-Bonnet coupling parameter $\alpha$ = 0.1 and 0.2 (left to right) with $\mu'=1$.}
\end{figure*}
The  Gibb's free energy (\ref{fe1}) reduces to the Gibb's free energy of EGB-AdS black hole \cite{cai,Ads,MS1} in the absence of charge $e=0$,
\begin{eqnarray}\label{fe2}
G_+&=&\frac{(D-2) V_{D-2}r_{+}^{D-3}}{16 \pi}\left[\left(1 + \frac{\tilde{\alpha}}{r_{+}^2}+\frac{r_{+}^2}{l^2}\right)-\left(\frac{r_+^2}{D-2}+\frac{2{\tilde \alpha}}{D-4}\right)\left[\frac{(D-3)r_{+}^2+(D-5){\tilde \alpha}+\frac{(D-1)r_+^4}{l^2}}{r_+^2(r_+^2+2{\tilde\alpha})}\right]\right]\nonumber\\
\end{eqnarray}
we recover the Gibb's free energy of the EGB black hole \cite{cai} when $e=0, \text{and}\, \Lambda=0$  and further we obtained the Gibb's free energy for $D$-dimensional Bardeen black hole \cite{sabir} in the limit $\alpha \to 0$  
\begin{eqnarray}\label{fe3}
G_+&=&\frac{(D-2) V_{D-2}(1+\frac{e^{D-2}}{r^{D-2}})^\frac{D-1}{D-2}r_{+}^{D-3}}{16 \pi}\Big[\left(1 +\frac{r_{+}^2}{l^2}\right)-\frac{(D-3)r_{+}^2-\frac{2e^{D-2}}{r_{+}^{D-2}}(r_{+}^2+\frac{D-1}{l^2}r_{+}^4)}{ r_{+}^2(1+\frac{e^{D-2}}{r_{+}^{D-2}})}\nonumber\\
&&\Big[(D-4)\frac{r_+^{D-2}}{e^{D-2}}H_1-(D-3)(D-4)H_3\Big]\Big].
\end{eqnarray}
The plot for Gibb's free energy with horizon radius in various dimensions has been shown in Fig. \ref{fig:fe}. Here, from Fig. \ref{fig:fe}, we noticed that the peaks of free energy increase and shift to the right as the value of  charge $e$ grows for given value of Gauss-Bonnet coupling parameter $\alpha$ and $\Lambda$. The peak also increase as the value of $\alpha$ grows. The behaviour of free energy \ref{fig:fe} suggests it is mostly positive for larger $r_+$. From Fig. \ref{fig:fe}, one can see
that Bardeen-EGB-AdS black hole more stable for smaller $r_+$.

\begin{figure*}
\begin{tabular}{c c c c}
\includegraphics[width=0.5\linewidth]{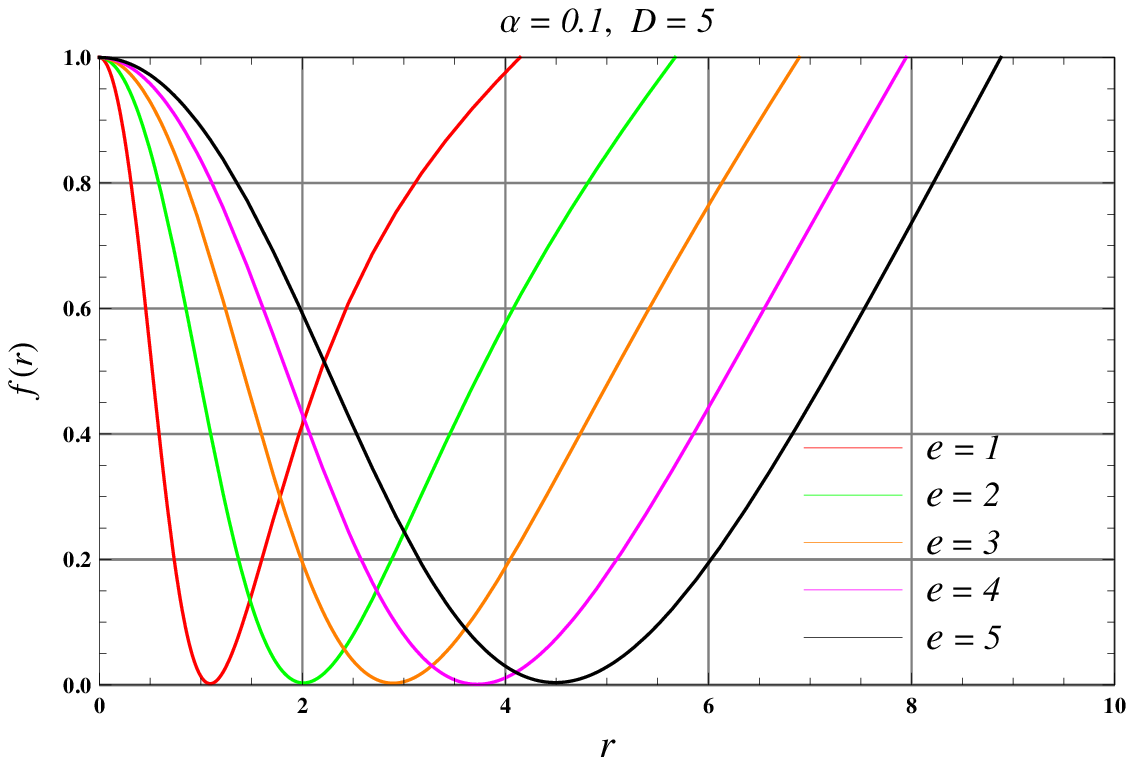}
\includegraphics[width=0.5\linewidth]{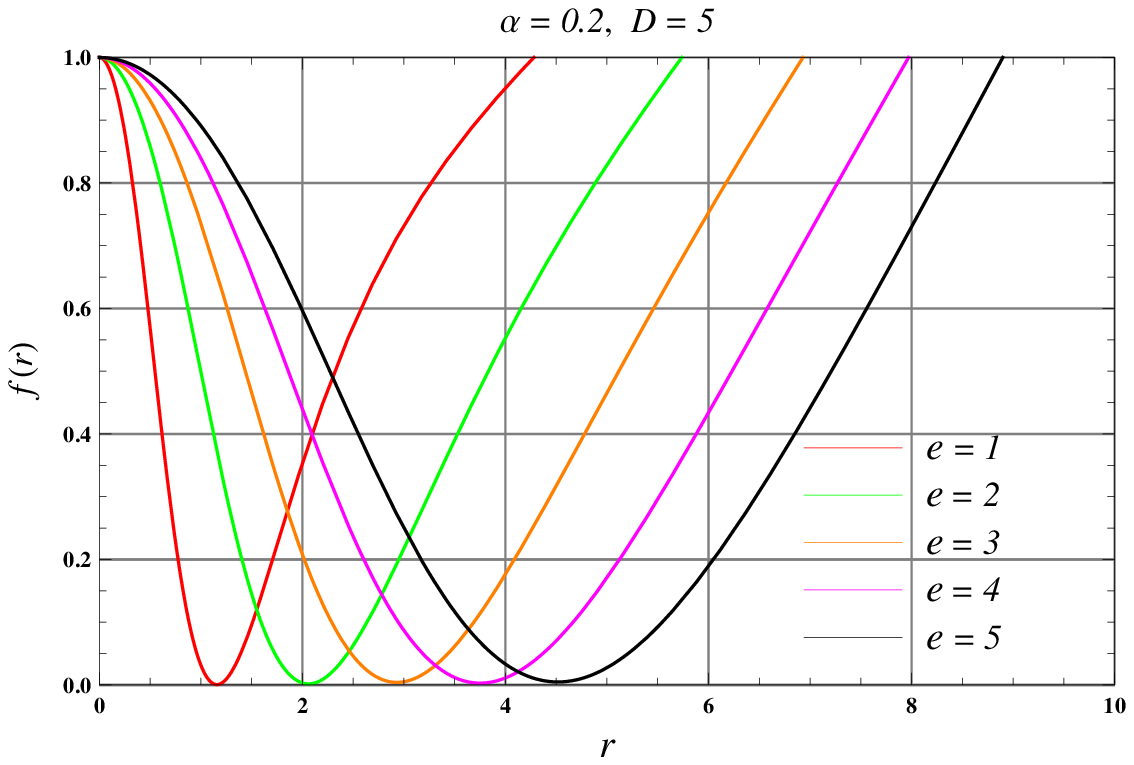}\\
\includegraphics[width=0.5\linewidth]{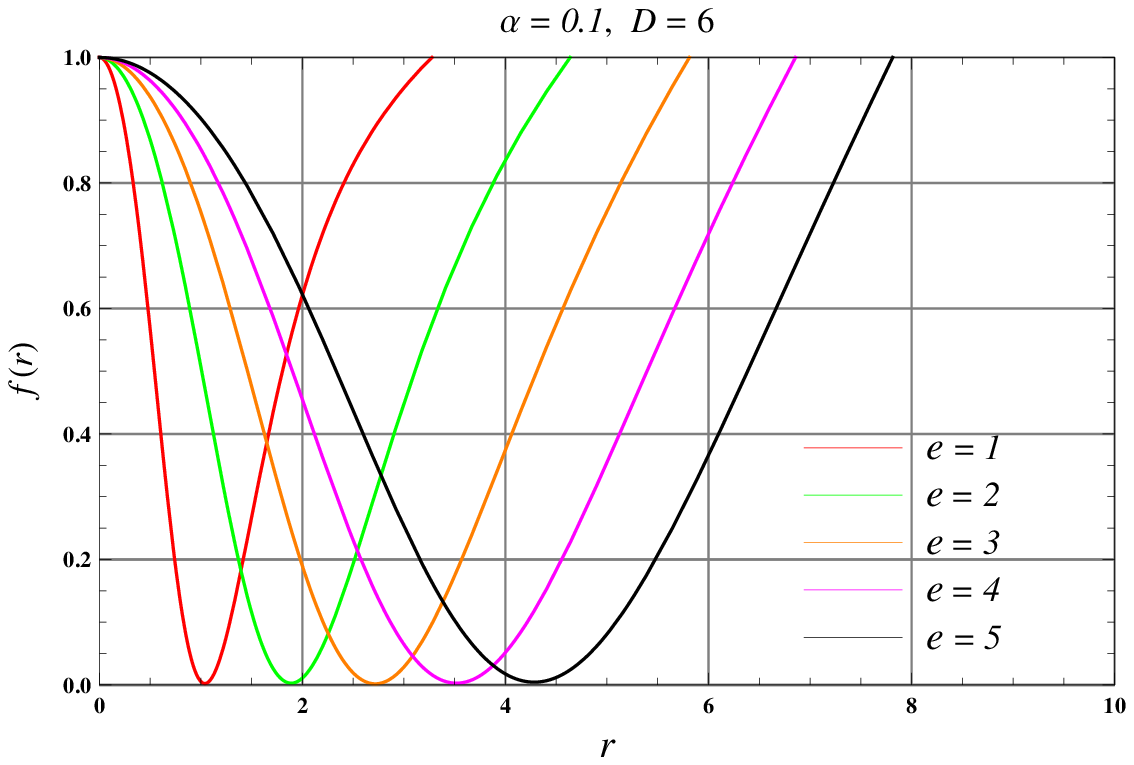}
\includegraphics[width=0.5\linewidth]{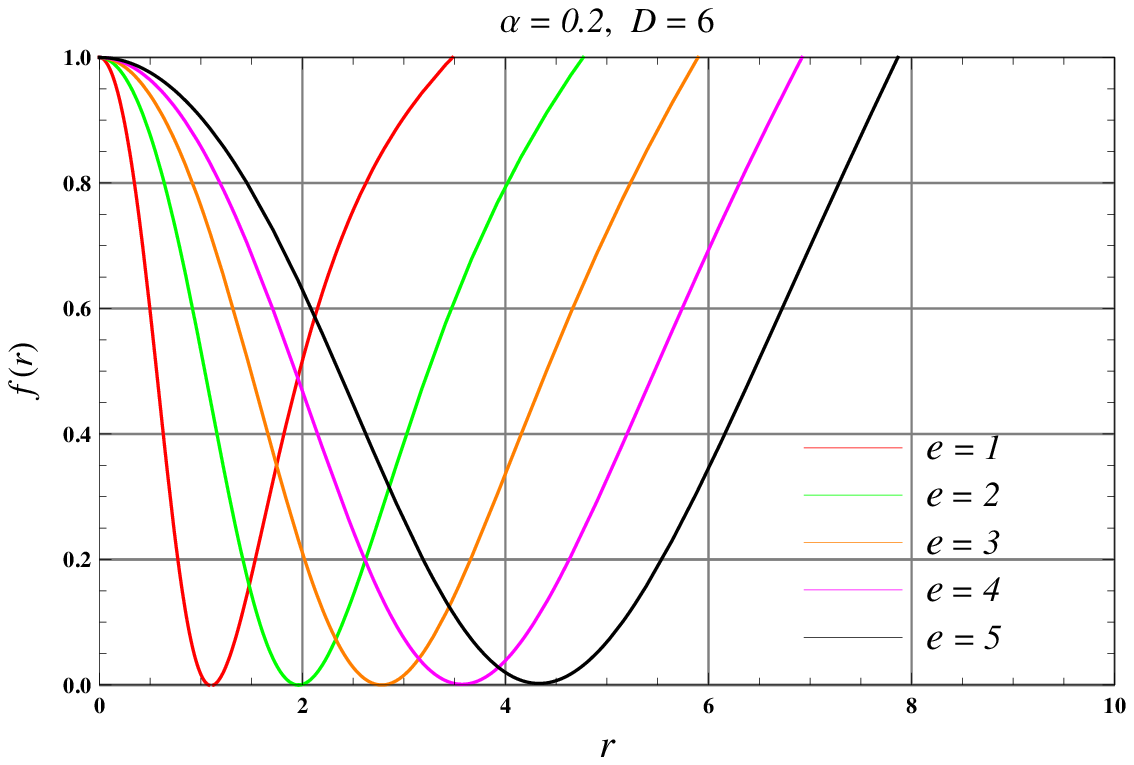}\\
\includegraphics[width=0.5\linewidth]{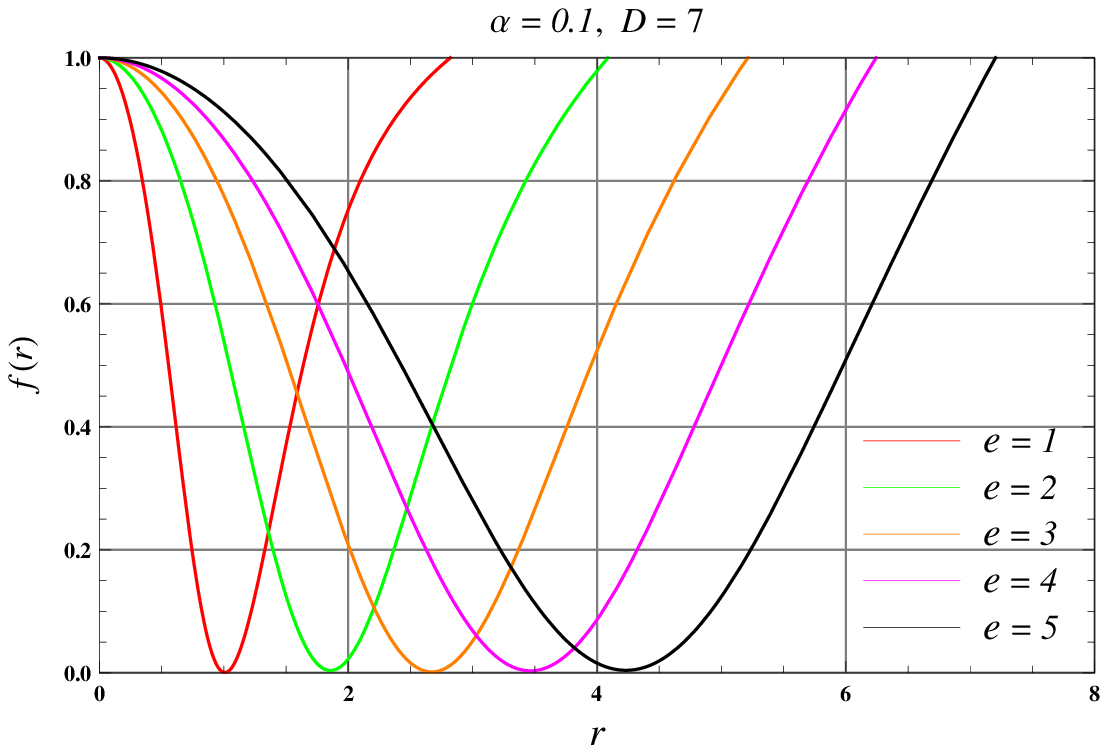}
\includegraphics[width=0.5\linewidth]{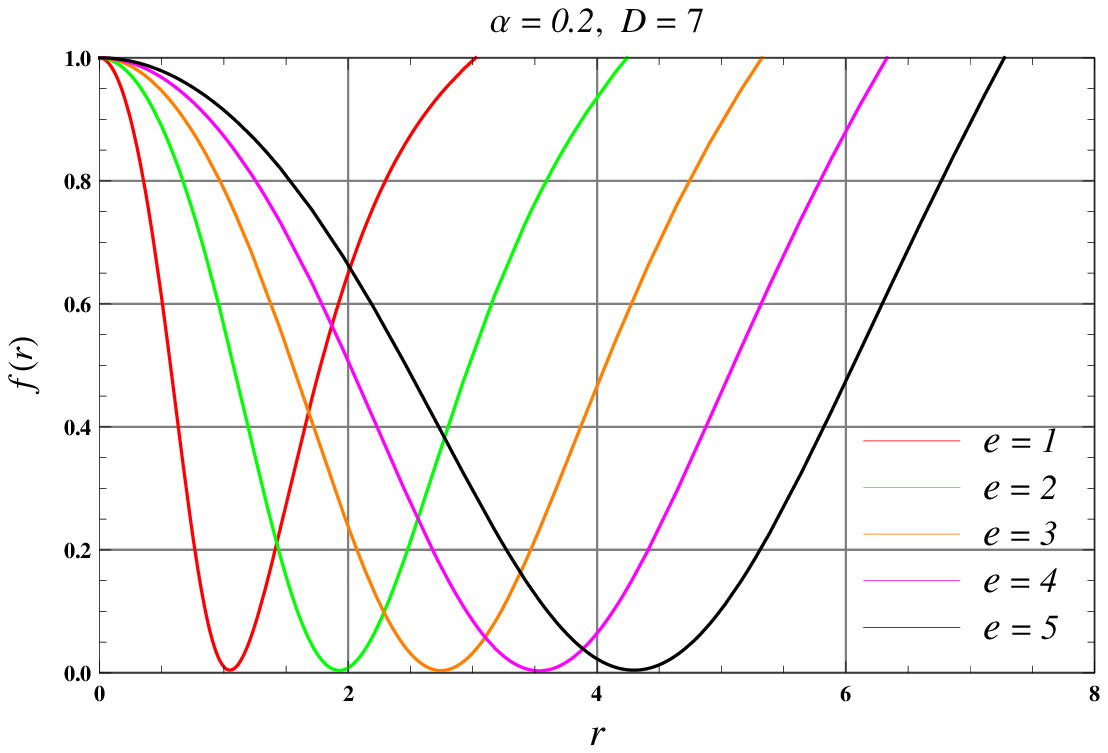}\\
\includegraphics[width=0.5\linewidth]{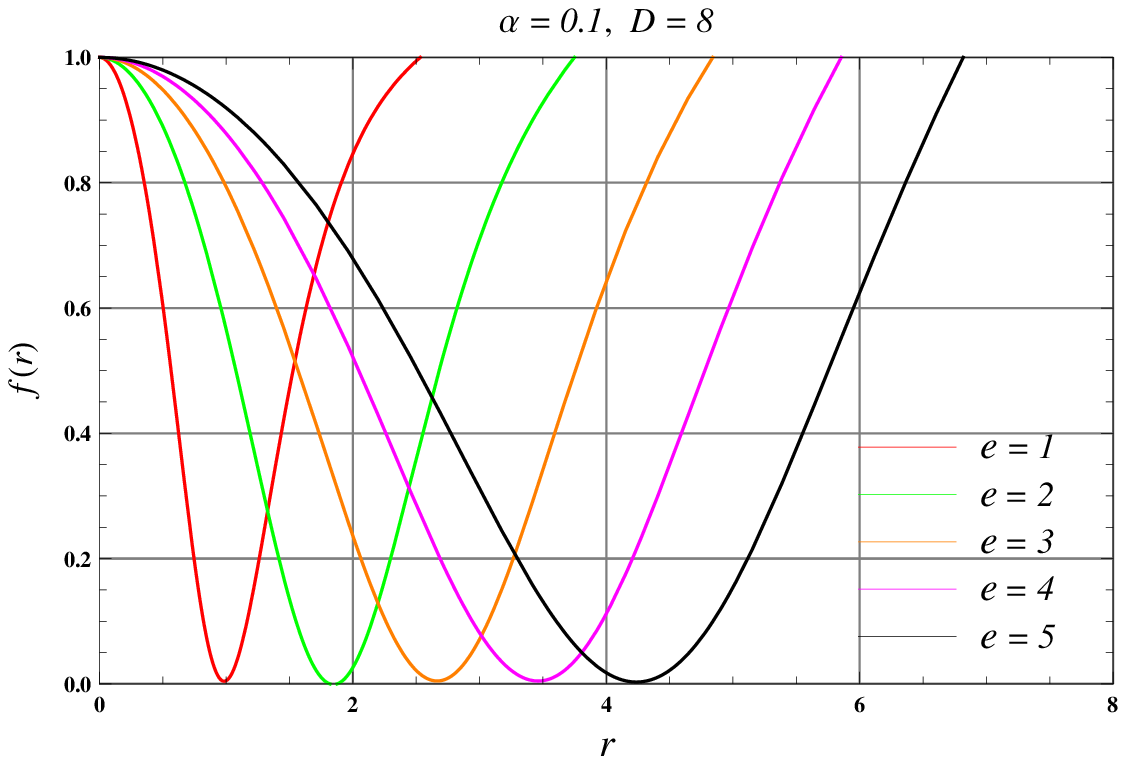}
\includegraphics[width=0.5\linewidth]{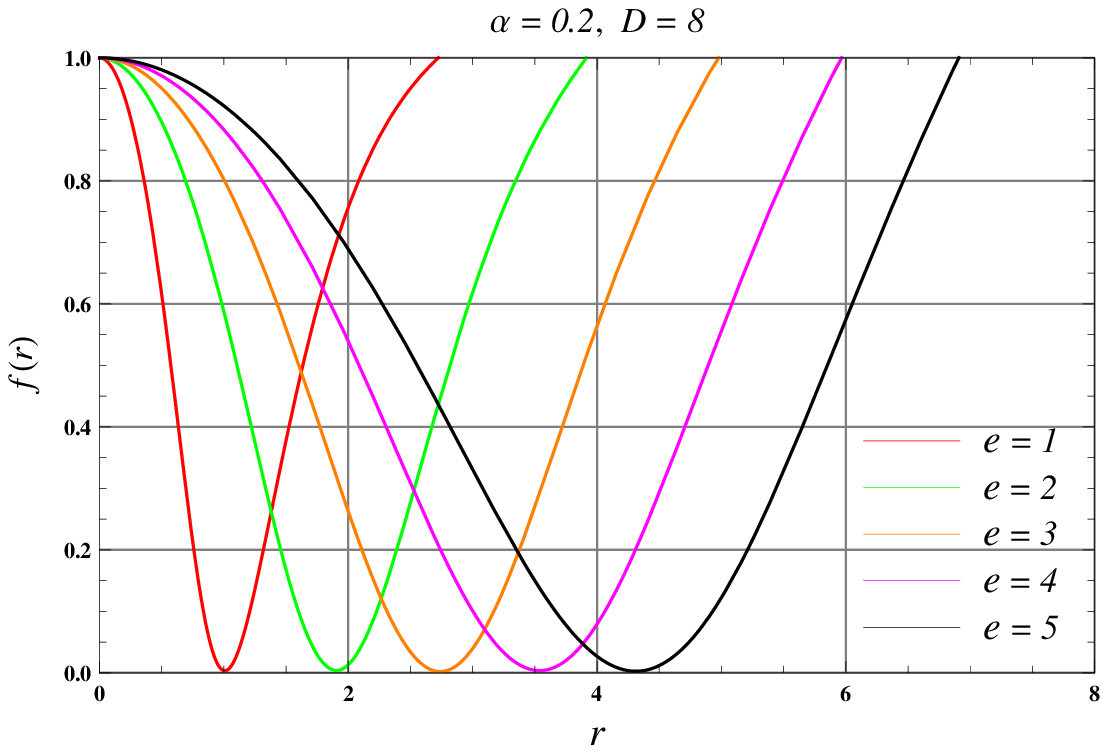}
\end{tabular}
\caption{\label{fig:rt} Plot of metric function $f(r)$ vs radius $r$ in various dimensions $D$ = 5, 6, 7, and 8 (top to bottom) for different values of  charge $e$ with Gauss-Bonnet coupling parameter $\alpha$ = 0.1 and 0.2 (left to right) with $\mu'=1$. }
\label{rem1}
\end{figure*}

The remnant of a black hole is a localized late stage of the black hole after the Hawking evaporation, which is either absolutely stable or long-lived \cite{PC}. It is very important to study the black hole remnant as it is a candidate to be the source of dark matter \cite{rem} as well to resolve the information loss paradox of black hole \cite{bhil}. We can get the radius $r_E$ of black hole remnant from $f^{\prime}(r)|_{r=r_E} = 0$. Here $r = r_E$ corresponds to the extremal black hole with the degenerate horizon. As it is very tedious to solve $f^{\prime}(r_E)=0$ analytically, so we tabulated the numerical results of remnant size and remnant term $\mu'$ in Table \ref{tr1} and Table \ref{tr2} in various dimensions for different values of  charge $e$ with Gauss-Bonnet coupling constant $\alpha=0.1$ and $\alpha=0.2$ respectively. The remnant mass $M_0$ can be calculated very easily by inserting the value of $\mu'_0$ in Eq. (\ref{a2}). In order to analyze the emitted features of Bardeen-EGB-AdS black hole, we plotted the metric function given in Eq. (\ref{sol:egb}) as a function of radius for extremal Bardeen-EGB-AdS black hole in Fig. \ref{fig:rt} for different values of $e$ and $\alpha$. From Fig. \ref{fig:rt}, we can say that at the minimal non zero mass $M_0$, there is a possibility of the extremal configuration with one degenerate event horizon. So, $M = M_0$ is the condition for having one degenerate event horizon and there will be no event horizon for $M<M_0$. 

\begin{table}[ht]
    \begin{center}
        \begin{tabular}{| l | l l  | l l  | l l  | l l  |  }
            \hline
  \multicolumn{1}{|c|}{Charge} &\multicolumn{2}{c|}{$D=5$}  &\multicolumn{2}{c|}{$D=6$}  &\multicolumn{2}{c|}{$D=7$} &\multicolumn{2}{c|}{$D=8$}\\
            \hline
            \,\,\,\,\,$e$ & $r_0$ &~~$\mu'_0$ & $r_0$ &~~$\mu'_0$  & $r_0$ & ~~$\mu'_0$ & $r_0$ & ~~$\mu'_0$\\
            \hline
            \,\,\,\,\,1 & 1.12  & ~~3     & 1.04 & ~~3.80   & 1.02 & ~~5.70       & 0.98     &~~ 6.70       \\
            \,\,\,\,\,2 & 2.00  & ~~10.90 & 1.92 & ~~22.37  & 1.86 & ~~47.93      & 1.82     & ~~106.9      \\
            \,\,\,\,\,3 & 2.89  & ~~25.10 & 2.71 & ~~72.20  & 2.69 & ~~215.5      & 2.67     & ~~660.4      \\
            \,\,\,\,\,4 & 3.76  & ~~46.80 & 3.48 & ~~173.83 & 3.46 & ~~668.2      & 3.45     & ~~2636      \\
            \,\,\,\,\,5 & 4.55  & ~~77.4  & 4.32 & ~~353.56 & 4.26 & ~~1669.56    & 4.24     & ~~8090       \\
            \hline
        \end{tabular}
    \end{center}
     \caption{The remnant size $r_0$ and the remnant mass term $\mu'_0$ for different values of parameter $e$ with Gauss-Bonnet coupling constant $\alpha=0.1$ with $l=10$.}
\label{tr1}
\end{table}
\begin{table}[h]
    \begin{center}        
        \begin{tabular}{| l | l l  | l l  | l l  | l l  |  }
            \hline
  \multicolumn{1}{|c|}{Charge} &\multicolumn{2}{c|}{$D=5$}  &\multicolumn{2}{c|}{$D=6$}  &\multicolumn{2}{c|}{$D=7$} &\multicolumn{2}{c|}{$D=8$}\\
            \hline
           \,\, \,\,$e$ & $r_0$ &~~$\mu'_0$ & $r_0$ &~~$\mu'_0$  & $r_0$ & ~~$\mu'_0$ & $r_0$& ~~$\mu'_0$ \\
            \hline
            \,\,\,\,\,1 & 1.14  & ~~3.41 & 1.12 & ~~5.12    & 1.05& ~~7.70       & 1.01   & ~~11.17       \\
            \,\,\,\,\,2 & 2.10  & ~~11.45 & 1.95 & ~~25.47   & 1.93 & ~~59.60      & 1.90  &~~143.7        \\
            \,\,\,\,\,3 & 2.93  & ~~25.60 & 2.78 & ~~77.30   & 2.75& ~~243.3        & 2.72  & ~~796.7      \\
            \,\,\,\,\,4 & 3.79  & ~~47.30 & 3.56 & ~~181.20  & 3.51 & ~~721.6      & 3.51   & ~~2972      \\
            \,\,\,\,\,5 & 4.57  & ~~77.96 & 4.34 & ~~363.58   & 4.32 & ~~1757       & 4.31   & ~~8775       \\
            \hline 
        \end{tabular}
        \caption{The remnant size $r_0$ and the remnant mass term $\mu'_0$ for different values of parameter $e$ with fixed value of Gauss-Bonnet coupling  constant $\alpha=0.2$ with $l=10$.}\label{tr2}
    \end{center}
\end{table}

\section{Conclusion}
The EGB gravity is a higher curvature generalization of general relativity which is also considered as quantum corrected gravity model and AdS black holes help us to understand the idea from quantum gravity as well as general relativity. Further, the holography beyond the AdS/CFT continues to exist in true quantum gravity that requires the inclusion of higher order curvature derivative term. Motivated by this, we studied exact static spherical $D$-dimensional Bardeen-EGB-AdS black holes and discuss their properties. Thus, we obtained an exact black hole in EGB gravity for a static and spherically symmetric $D$-dimensional AdS spacetime with energy-momentum given by a nonlinear electrodynamics. The solution is characterized by analyzing horizons which could be at the most two. 

Later, we compute exact expressions for Hawking temperature, entropy, heat capacity and free energy associated with the black holes, also demonstrate that Hawking-Page transition is achievable. We perform a detailed analysis of the thermodynamical specific heat with focus on the local and global stability. It turns out that heat capacity can be negative or positive depending on the choice of parameters $e$ and $\alpha$, which further, respectively, tells us that the black hole is unstable or stable.

Indeed, the phase transition of black hole is characterized by the divergence of its specific heat at a critical horizon radius $r^{C}_+$ which is varying with the spacetime dimension $D$ and parameter $\alpha$. The black holes are thermodynamically stable with a positive heat capacity for the range $r_1<r_+<r^{C}_+$ and unstable for $r_1>r_+> r^{C}_+$ (cf. Fig. \ref{fig:st}). We discussed the black hole remnant and tabulated the numerical values of black hole remnant size and mass. The results presented here are the generalization of the previous discussions and in the appropriate limits, go over to AdS-EGB black holes and EGB black holes. The possibility of a further generalization of these results to Lovelock gravity is an interesting problem for future research.
 
\acknowledgements
 D.V.S. acknowledges the University Grant Commission, India, for financial support through the D. S. Kothari Post Doctoral Fellowship (Grant No.: BSR/2015-16/PH/0014). S.G.G. would like to thanks SERB-DST Research Project Grant No.SB/S2/HEP-008/2014 and DST INDO-SA bilateral project DST/INT/South Africa/P06/2016 and also to IUCAA, Pune for the hospitality while this work was being done.


\begin{thebibliography}{99}       
\bibitem{1} R. Penrose,  Phys. Rev. Lett. {\bf 14}, 57 (1965); S. W. Hawking,  Proc. Roy. Soc. Lond. A {\bf 300}, 187 (1967); S. W. Hawking and R. Penrose,  Proc. Roy. Soc. Lond. A 3{\bf 14}, 529 (1970).

\bibitem{rp} R. Penrose, Riv. Nuovo Cimento {\bf 1}, 252 (1969); in General Relativity, an Einstein
	Centenary Volume, edited by S. W. Hawking and W. Israel (Cambridge University
	Press, Cambridge, England, 1979).


	\bibitem{Sakharov:1966}
	A.D.~Sakharov,
	Sov.\ Phys.\ JETP, {\bf 22}, 241 (1966).
	
	\bibitem{Gliner:1966}
	E.B.~Gliner,
	Sov.\ Phys.\ JETP, {\bf 22}, 378 (1966).
	
	\bibitem{Bardeen:1968}
	J.~Bardeen,
	in {\it Proceedings of GR5}, Tiflis, U.S.S.R. (1968).
\bibitem{ansoldi}
S. Ansoldi, arXiv 0802.0330 [gr-qc].
\bibitem{ABG99}E. Ayon-Beato and A. Garcia, Phys. Lett. B {\bf 493}, 149 (2000); E.~Ayon-Beato and A.~Garcia, Gen.\ Rel.\ Grav.\  {\bf 31}, 629 (1999);  E. Ayon-Beato and A. Garcia, Gen. Rel. Grav. {\bf 37}, 635 (2005); E.~Ayon-Beato and A.~Garcia, Phys.\ Rev.\ Lett.\  {\bf 80}, 5056 (1998).

\bibitem{8} L.~Xiang, Y.~Ling and Y.~G.~Shen, Int.\ J.\ Mod.\ Phys.\ D {\bf 22}, 1342016 (2013);  H.~Culetu,  Int.\ J.\ Theor.\ Phys.\  {\bf 54},  2855 (2015); L.~Balart and E.~C.~Vagenas,  Phys.\ Lett.\ B {\bf 730}, 14 (2014);  L.~Balart and E.~C.~Vagenas,  Phys.\ Rev.\ D {\bf 90}, 124045 (2014).

\bibitem{Fernando:2016ksb} 
  S.~Fernando,
  Int.\ J.\ Mod.\ Phys.\ D {\bf 26}, 1750071 (2017).
\bibitem{Bambi}
 C.~Bambi and L.~Modesto,
  Phys.\ Lett.\ B {\bf 721},  329 (2013).
\bibitem{Sharif:2011ja} 
  M.~Sharif and W.~Javed,
  Can.\ J.\ Phys.\  {\bf 89}, 1027 (2011).
\bibitem{12} 
  K.~Ghaderi and B.~Malakolkalami, Grav.\ Cosmol.\  {\bf 24}, 61 (2018);  N.~Bretón and S.~E.~Perez Bergliaffa, AIP Conf.\ Proc.\  {\bf 1577}, 112 (2014);  J.~Man and H.~Cheng, Gen.\ Rel.\ Grav.\  {\bf 46}, 1660 (2014);. 
\bibitem{Stuchlik:2014qja} 
  Z.~Stuchlík and J.~Schee,
  Int.\ J.\ Mod.\ Phys.\ D {\bf 24},  1550020 (2014).
\bibitem{14} 
  S.~Fernando and J.~Correa,
  Phys.\ Rev.\ D {\bf 86}, 064039 (2012);  C.~F.~B.~Macedo, L.~C.~B.~Crispino and E.~S.~de Oliveira,
  Int.\ J.\ Mod.\ Phys.\ D {\bf 25},  1641008 (2016);  S.~C.~Ulhoa,
  Braz.\ J.\ Phys.\  {\bf 44}, 380 (2014);  W.~Wahlang, P.~A.~Jeena and S.~Chakrabarti,
  Int.\ J.\ Mod.\ Phys.\ D {\bf 26}, 1750160 (2017);  M.~Saleh, B.~B.~Thomas and T.~C.~Kofane,
  Eur.\ Phys.\ J.\ C {\bf 78}, 325 (2018).

\bibitem{15} 
  D.~V.~Singh and N.~K.~Singh,
  Ann. Phys.\  {\bf 383}, 600 (2017);  H.~Huang, M.~Jiang, J.~Chen and Y.~Wang,
  Gen.\ Rel.\ Grav.\  {\bf 47},  8 (2015).
\bibitem{Mehdipour:2016rtd} 
  S.~H.~Mehdipour and M.~H.~Ahmadi,
  Astrophys.\ Space Sci.\  {\bf 361}, 314 (2016).
\bibitem{Bambi:2014nta} 
  C.~Bambi,
  Phys.\ Lett.\ B {\bf 730}, 59 (2014).

\bibitem{Ghosh:2015pra} 
  S.~G.~Ghosh and M.~Amir,
  Eur.\ Phys.\ J.\ C {\bf 75}, 553 (2015).
  \bibitem{sabir}
 Md. Sabir Ali and S. G. Ghosh, $d$-dimensional Bardeen black hole and thermodynamics.
\bibitem{gross} D. Gross and E. Witten,  Nucl. Phys.
B {\bf 277}, 1 (1986).
\bibitem{5}
C. Lanczos, Ann. Math. {\bf 39}, 842 (1938).
\bibitem{6}
 D. Lovelock, J. Math. Phys. (N.Y.) {\bf 12}, 498 (1971).
\bibitem{BD} D. G. Boulware and S. Deser, Phys. Rev. Lett. {\bf 55}, 2656 (1985).
\bibitem{cai} R.G. Cai, Phys. Rev. D {\bf 65}, 084014 (2002).
\bibitem{26} J. T. Wheeler, Nucl. Phys. B {\bf 268}, 737 (1986); J. T. Wheeler,  Nucl. Phys. B {\bf 273},  732 (1986); S. G. Ghosh and D. W. Deshkar, Phys. Rev. D { \bf 77}, 047504 (2008); S. G. Ghosh, Phys. Lett. B {\bf 704}, 5 (2011); R. C. Myers and J. Z. Simon, Phys. Rev. D {\bf 38}, 2434 (1988);  M. H. Dehghani and R. B. Mann,  Phys. Rev. D {\bf 72}, 124006 (2005); S. G. Ghosh and S.D. Maharaj, Phys. Rev. D, {\bf 89}, 084027 (2014); H. Maeda and N. Dadhich,  Phys. Rev. D {\bf 75} 044007 (2007);  S.~G.~Ghosh, D.~V.~Singh and S.~D.~Maharaj,
Phys.\ Rev.\ D {\bf 97}, 104050 (2018);  G. Kofinas and R. Olea, Phys. Rev. D {\bf 74}, 084035 (2006); S.~G.~Ghosh,
Classical Quantum Gravity\  {\bf 35}, 085008 (2018);  D.~V.~Singh, M.~S.~Ali and S.~G.~Ghosh, Int.\ J.\ Mod.\ Phys.\ D {\bf 27}, 1850108 (2018).
\bibitem {sus} S. G. Ghosh, U. Papnoi, and S. D. Maharaj, Phys. Rev. D {\bf 90}, 044068 (2014). 
\bibitem{27} T. Torii and H. Maeda, Phys. Rev. D {\bf 71}, 124002 (2005).
\bibitem{Neu} Y. M. Cho and I. P. Neupane, Phys. Rev. D {\bf 66}, 024044 (2002); I. P. Neupane, Phys. Rev. D {\bf 69}, 084011 (2004).
\bibitem {MS1} M. H. Dehghani and S. H. Hendi, Int. J. Mod. Phys. D {\bf 16}, 1829 (2007).
\bibitem{djg}
D. J. Gross and J. H Sloan, Nucl. Phys. B {\bf 291}, 41 (1987).
\bibitem{bento}
 M. C. Bento  and O. Bertolami, Phys. Lett. B {\bf 368}, 198 (1996). 


\bibitem{Hendi1}
S. H. Hendi, S. Panahiyan and B. Eslam Panah, J. High Energy Phys. {\bf 01}, 129 (2016).

\bibitem {Ads} N. Deruelle and L. Farina-Busto, Phys. Rev. D {\bf 41}, 3696 (1990);  M. H. Dehghani, Phys. Rev. D {\bf 69}, 064024 (2004);  N. Deruelle, J. Katz, and S. Ogushi, Classical Quantum Gravity {\bf 21}, 1971 (2004); M. H. Dehghani and S. H. Hendi, Phys. Rev. D {\bf 73}, 084021 (2006); A. Padilla, Classical Quantum Gravity {\bf 20}, 3129 (2003);  M. H. Dehghani, G. H. Bordbar, and M. Shamirzaie, Phys. Rev. D {\bf 74}, 064023 (2006).
\bibitem {st}  F. R. Tangherlini, Nuovo Cim. {\bf 27}, 636 (1963).
\bibitem {hp} S. W. Hawking, D. N. Page, Commun. Math. Phys. {\bf 87}, 577 (1983).

\bibitem {Wij} C.~ Sahabandu, P.~ Suranyi, C.~Vaz, and L.~ C.~ R.~ Wijewardhana, Phys. Rev. D {\bf 73}, 044009 (2006).
\bibitem {PK} P. Kanti, Lect. Notes Phys., {\bf 769}, 387 423, (2009).

\bibitem {thermo} J. M. Bardeen, B. Carter, and S.W. Hawking, Commun. Math. Phys. {\bf 31}, 161-170 (1973).
\bibitem {jdb} J.~ D.~Bekenstein, Phys. Rev. D {\bf 7}, 2333 (1973).

\bibitem{Herscovich}
E. Herscovich, M. G. Richarte, Phys. Lett. B {\bf 689}, 192–200 (2010). 
\bibitem {PC} P. Chen, Y. C. Ong, D. H. Yeom, Phys. Rept. {\bf 603} (2015).  
\bibitem {rem} J. H.~MacGibbon, Nature {\bf 329}, 308 (1987).
\bibitem {bhil} J.~Preskill, arXiv: 9209058 [hep-th].
  



\end{thebibliography}
\end{document}